\documentclass[twocolumn]{aastex631}

\usepackage{threeparttable}
\usepackage{parskip}
\usepackage{float}
\usepackage{hyperref}
\usepackage{graphicx}

\usepackage{booktabs}

\usepackage{amsthm}
\usepackage{amsmath}

\usepackage{soul}

\newcommand{\red}[1]{\textcolor{red}{#1}}

\begin{document}

\title{Insights into the Long-Term  Flaring Events of Blazar PKS 0805-07: A Multi-Wavelength Analysis over the 2009-2023 Period}

\author{Sikandar Akbar}
\affiliation{Department of Physics, University of Kashmir, Srinagar 190006, India.}

\author{Zahir Shah}
\affiliation{Department of Physics, Central University of Kashmir, Ganderbal 191201, India.}

\author{Ranjeev Misra}
\affiliation{Inter-University Center for Astronomy and Astrophysics, Post Bag 4, Ganeshkhind, Pune,411007 India.}

\author{Naseer Iqbal}
\altaffiliation{Department of Physics, University of Kashmir, Srinagar 190006, India.}

\begin{abstract}
We conducted a comprehensive temporal and spectral study of the FSRQ PKS 0805-07 by using the broadband observations from the Fermi-LAT and Swift-XRT/UVOT instruments over the period MJD 54684-60264. 
The 3-day binned $\gamma$-ray  light curve during the active state, revealed eleven distinct peak structures with the maximum integral flux (E $>$ 100 MeV) reached $(1.56\pm 0.16)\times10^{-6}\, \text{photons cm}^{-2}\, \text{s}^{-1}$ on MJD 59904.5. The shortest observed $\gamma$-ray variability was $2.80\pm 0.77$ days. A correlation analysis between the $\gamma$-ray spectral index and flux indicated the typical trend of hardening when the source is brighter, commonly observed in blazars.
We identified a lag of 121 (+27.21, -3.51) days in the spectral index relative to the flux, within the time interval MJD 59582 to 60112. The Anderson-Darling test and histogram-fit rejected the normality of the $\gamma$-ray flux distribution, and instead suggest a log-normal distribution.
To gain insight into the underlying physical processes, we extracted broadband spectra from different time periods in the light curve. The spectral energy distribution during various flux states were well-reproduced using synchrotron, synchrotron-self-Compton, and external-Compton emissions from a broken power-law electron distribution. The seed photons required for the external Compton process are from IR region.  A comparison of the best-fit physical parameters indicated that the variations in different flux states were primarily associated with an increase in the bulk Lorentz factor and magnetic field from low to high flux states.
\end{abstract}



\section{Introduction}
\label{introduction}

PKS 0805-07 is a  flat-spectrum radio quasar (FSRQ), located at a high redshift z = 1.837  \citep{1988ApJ...327..561W}. FSRQ is a subclass of most distinctive and powerful Active Galactic Nuclei (AGNs) called Blazars. The majority of extra-galactic entities observed in the $\gamma$-ray sky, as detected by instruments like the Fermi Gamma-ray Space Telescope \citep{2010ApJS..188..405A}, are represented by blazars. Blazars exhibits exceptional observational properties which are attributed to the presence of a powerful non-thermal relativistic jet of  plasma pointing toward the observer, originating from the vicinity of a central supermassive black hole  \citep{1995PASP..107..803U}. These include rapid flux and spectrum variability across all wavelength bands with timescales ranging from a few minutes to years , high and variable polarization, high energetic gamma-ray emissions and apparent superluminal motion as documented in numerous studies (e.g.,\citet{2007ApJ...664L..71A, 2016ApJ...824L..20A, 2006A&A...453..817V, Fan_2021,Xiao_2022, 10.1093/mnras/stad3574,2022ApJ...938L...7D, 2022Natur.609..265J, 2011IAUS..275..164F,2023arXiv230606863Q}). 

PKS 0805-07 is listed in the Fermi 4FGL catalog under the identifier 4FGL J0808.2-0751 \citep{2020ApJS..247...33A}. This source is well known for its exceptionally high apparent jet motions \citep{2019ApJ...874...43L}. The active state of the source has been consistently reported in multiple astronomical telegrams (e.g. references). \citet{2009ATel.2136....1C} reported a high state with a $\gamma$-ray flux (E$>$100MeV) of $1.60 \pm 0.33 \times 10^{-6}\,\text{photons\, cm}^{-2} \,\text{s}^{-1}$ on July 22, 2009, almost a factor of 2 higher than the flux reported by \cite{2009ATel.2048....1C}.  
On October 11, 2022, this source displayed an elevated state of $\gamma$-ray emissions, with a daily averaged flux (E$>$100MeV) of approximately $(1.4 \pm 0.2) \times 110^{-6}\,\text{photons\, cm}^{-2} \,\text{s}^{-1}$. This marks a 16-fold increase compared to the average flux reported in the third release of the Fourth Fermi-LAT catalog \citep{2022ApJS..260...53A}. %(4FGL-DR3, Abdollahi et al. 2022, ApJS, 260, 53).
 The corresponding photon index was approximately $1.97 \pm 0.09$, notably lower than the 4FGL-DR3 value of $2.23 \pm 0.01$. The spectral hardening persisted on October 12, 2022, with the source detected at a flux of approximately $(1.2 \pm 0.2) \times 10^{-6}\,\text{photons\, cm}^{-2} \,\text{s}^{-1}$  and a spectral index of approximately $1.98 \pm 0.10$ \citep{2022ATel15676....1L}. This event represented the second most significant $\gamma$-ray flare observed by LAT from this source, following the flux reported on July 22, 2009 \cite{2009ATel.2136....1C}.

The Australia Telescope Compact Array (ATCA) has been observing PKS\,0805-07 across multiple frequencies, ranging from 2.1 GHz to 45 GHz, since 2006. The majority of these observations were conducted as part of the Observatory-led Active Galactic Nuclei (AGN) monitoring program (C007). The latest measurements are derived from the TANAMI/C1730 program, targeting $\gamma$-ray bright blazars in the southern sky. In addition, PKS 0805-07 has been part of the Metsahovi AGN monitoring program since mid-2002, and from 2009 onward, it has been consistently monitored at 37 GHz with a high cadence.

Starting in early 2021, the high-frequency emission from PKS 0805-07 has shown a significant increase above its long-term average state. 
The most recent measurements, recorded on September 25, 2021, and October 12, 2021 indicate historically high flux densities for this source, with values of 4.0 $\pm$ 0.2 Jy and 4.1 $\pm$ 0.2 Jy, respectively, in both the ATCA and Metsahovi monitoring programs \citep{2022ATel15692....1E}. Furthermore, the GRID detector on the AGILE satellite has detected a prominent $\gamma$-ray flare from PKS 0805-07. A preliminary multi-source likelihood analysis indicates a 6 sigma detection, with a $\gamma$-ray flux (F($>$100 \, MeV)) of $(2.7 \pm 0.8) \times 10^{-6} \, photons\, cm^{-2} s^{-1}$ during the period of November 17-19, 2022  \citep{2022ATel15768....1B}. This marks a substantial increase compared to the average flux observed in the preceding seven days.

The spectral energy distribution (SED) of blazars typically displays a distinctive double-peaked pattern. The theoretical models used to interpret the observed broadband SED of blazars mainly fall into two categories: leptonic and hadronic origins. The low-energy component, peaking at optical/UV/X-ray energies, is attributed to Doppler-boosted synchrotron emission. The high-energy component, peaking in the $\gamma$-ray range, is often explained by leptonic process via Inverse Compton (IC;  \citet{2013ApJ...768...54B, 2017MNRAS.470.3283S}) and
hadronic emission processes through the proton-synchrotron process or proton-photon interactions \citep{2003APh....18..593M, 2001APh....15..121M, 1993A&A...269...67M}. These interactions can give rise to $\gamma$-ray photons and high-energy neutrinos. Inverse Compton (IC) involves primarily two processes. One is synchrotron self Compton (SSC), where relativistic particles in the jet cause the up-scattering of synchrotron photons \citep{1974ApJ...188..353J, 1992ApJ...397L...5M, 1993ApJ...407...65G}. Another is the external Compton (EC) process, where relativistic particles in the jet up-scatter photons external to the jet \citep{1992A&A...256L..27D, 1994ApJ...421..153S, 2000ApJ...545..107B, 2017MNRAS.470.3283S}.

In a recent study conducted by \citet{2017MNRAS.471.3036P}, the authors identified indications of periodic behaviors in four blazars: 4C + 01.28, S5 0716+71, PKS 0805 -07, and PKS 2052-47 apart from the previously asserted quasi-periodic $\gamma$-ray blazars. Among these, three sources including PKS 0805-07 are situated at elevated redshifts and emerge as potential candidates for binary systems of supermassive black holes (SMBHs).

The identification of neutrino events from blazars TXS 0506+056 associated with the high-energy neutrino IC-170922A \citep{2018Sci...361.1378I}, has sparked increased interest in hybrid models, particularly lepto-hadronic models. These models aim to enhance our understanding of the broadband Spectral Energy Distribution (SED) of blazars \citep{2018ApJ...863L..10A,2018ApJ...866..109S,2022MNRAS.509.2102G,2018ApJ...864...84K}.

Despite the extensive list of high-flux detections, a comprehensive long-term investigation of the source has been lacking until now. Our study is the first to conduct comprehensive broadband temporal and spectral analyses of PKS\,0805-07 using multi-wavelength observations from Fermi-LAT and Swift-XRT/UVOT. This allows us to examine its variability in unprecedented detail.  Additionally, the motivation for studying PKS\,0805-07 lies in its potential to serve as a case study for testing the theoretical models. In our work, we used the broadband observations of the source in different flux states, to check the consistancy of our convolved one zone laptonic model.  The model is incorporated as local convolution model in XSPEC and outputs broadband spectrum for any input particle distribution. This model is tested previously for the BL\,Lac source \citep{ 2024MNRAS.527.5140S}. Our work aims to highlight the importance of individual case studies in astrophysics, demonstrating  the importance of continuous observation and analysis of individual blazars, as even well-studied objects can reveal important features under different observational conditions.
The structure of this paper is as follows: Section \S2 provides details on the multi-wavelength data and the procedures for data analysis, while Section \S3 presents the broadband spectral and temporal findings of the source. The concluding section, \S4, summarizes and discusses the results.

\section{Observations and Data reduction} \label{sec:data_ana}
\subsection{Fermi-LAT}

Fermi-LAT is a high-energy space-based telescope aboard the Fermi Gamma-ray Space Telescope (previously GLAST). It was launched by NASA in 2008 and  has a wide field of view of approximately 2.3 Sr. Operating primarily in scanning mode, it surveys the entire sky in the energy range of about 20 MeV to 500 GeV every three hours \citep{2009ApJ...697.1071A}. This study focused on retrieving $\gamma$-ray data from Fermi-LAT for PKS\,0805-07 during the time period MJD 54684–60264.  For temporal and spectral analysis, the data were processed using Fermitools (formerly Science Tools) version 2.2.0, available on an Anaconda Cloud channel maintained by the Fermi Science Support Center (FSSC). Standard analysis procedures outlined in the Fermi-LAT documentation were followed for data reduction \footnote{http:// fermi.gsfc.nasa.gov/ ssc/ data/ analysis/}. P8R3 events were extracted from a 15-degree region of interest centered at the source location, with a focus on including high-probability photon events of the SOURCE class (evclass=128, evtype=3). To mitigate background $\gamma$-rays from the Earth limb, a zenith angle cut of 90 degrees was applied to the data. In the spectral analysis, we focused on photons within the energy range of 0.1–300 GeV. Additionally,  we utilised  FERMIPY, v1.0.1 \cite{2017ICRC...35..824W}, for the analysis. The galactic diffuse emission component was characterised using the $gll\_iem\_v07.fits$ model, and the isotropic emission component was represented by $iso\_P8R3\_SOURCE\_V 3\_v1.txt$. Our work employed the post-launch instrument response function $P8R3\_SOURCE\_V3$. All the sources located within a 25  degree region of interest (ROI) centered on the source position in the 4FGL catalog were incorporated into the XML model file. For the spectral analysis,  we kept the normalisation of  the sources lying within 10 degree of ROI as free parameters. In addition to normalisation,  the spectral parameters (alpha and beta) of  PKS\,0805-07 were also kept free.  All other parameters of the sources were kept freezed to their 4FGL catalog value. We generated  SED’s for various activity periods. We generated 3-day, 7-day, 15-day, 1-month, 3-month, and 6-month binned $\gamma$-ray light curves  for the temporal study.
 \\

\subsection{Swift-XRT}

The X-ray data utilized in this study were acquired through the Swift-XRT instrument aboard the Neil Gehrels Swift Observatory \citep{2004ApJ...611.1005G}. The Swift-XRT light curve was derived, with each observation ID corresponding to a distinct point in the X-ray light curve. To process the X-ray data collected in photon-counting mode, we utilized the XRTDAS V3.0.0 software package, which is an integral part of the HEASoft package (version 6.29). Following the guidelines provided in the Swift analysis thread page, we used the standard XRTPIPELINE (version: 0.13.6) to generate level 2 cleaned event files. Source events for spectral analysis were selected from a circular region with a radius of 50 arcsec, while the  background spectra were selected from a circular region with a radius of 100 arcsec outside the source region. The XIMAGE tool was employed to integrate exposure maps, and the xrtmkarf task was utilized to generate auxiliary response files. Using the GRPPHA task, source spectra were binned to ensure that each bin contained a minimum of 20 counts.  For the spectral analysis, XSPEC version 12.12.0  was employed \citep{1996ASPC..101...17A}. In order to account for the absorption effects due to neutral hydrogen, we used the Tbabs model. The neutral hydrogen column density (nH) value were fixed at a constant value of $9.60\times 10^{20}$ $cm^{-2}$ \citep{2005A&A...440..775K}.

\subsection{Swift-UVOT}
Besides the X-ray information, Swift also gives us data in the Optical/UV range using the Swift-UVOT telescope \citep{2005SSRv..120...95R}. This telescope observes in optical and UV using different filters like v, b, u, w1, m2, and w2 \citep{2008MNRAS.383..627P,2010MNRAS.406.1687B}.
We obtained the PKS 0805-07 data from the HEASARC Archive. We used the HEASoft package (v6.29) to extract useful scientific products from the data. To process the images, we used a package called uvotsource from the HEASoft package (v6.26.1). We added multiple images taken with different filters using the uvotimsum tool.
We extracted  the source counts by focusing on a circular area with a radius of $5''$ centered at the source location. The background region was taken from a  nearby source free circular area of radius of $10''$. 
To make our observations more accurate, we adjusted for the impact of dust in our galaxy. This involved correcting for Galactic extinction using specific values $E(B - V)=0.3711$  and $ R_{V} = A_{V}/E(B - V) =3.1$
\citep{2007ApJ...663..320F}.

\section{Results}
\subsection{Temporal Analysis}

To explore the temporal behavior of the source, we generated 3-day, 7-day, 15-day, 1-month, 3-month, and 6-month binned long-term $\gamma$-ray light curves. For the temporal analysis, we retained  the time bins which have detection significance (TS$\ge$4). The flux in the light curves are  integrated over the energy range of 0.1–100 GeV. This dataset spans the period from MJD 54684 to 60264. The Table \ref{table:peak} presents the maximum $\gamma$-ray flux, along with the associated spectral index,  time and TS value, for various time-binned $\gamma$-ray light curves.

\begin{table*}
\caption{Col. 1: denotes the time bin, 2: denotes the time (MJD) corresponding to maximum flux, 3: represents the maximum flux (measured in units of $10^{-6}\, \text{photons\, cm}^{-2} \,\text{s}^{-1}$), 4 and 5: display the spectral index and TS Value, 6 and 7: represent the averaged flux ($F_{av}$ , in units of $10^{-7}\, \text{photons\, cm}^{-2} \,\text{s}^{-1}$) between MJD 54684\textendash{}59370 and MJD 59370\textendash{}59965, respectively.}

\label{table:peak}
\centering
\begin{tabular}{lcccccc}
Time Bin &  Time &  Flux &  Index  &  TS & $F_{av}$ (54684\textendash{}59370) &$F_{av}$ (59370\textendash{}59965)\\
\hline

3-day    &   59904.49  & $1.56\pm 0.16$  &   $2.13\pm 0.08$ & 456.63& $0.49\pm 0.01$ & $2.10\pm 0.05$\\

7-day    &   59901.49  &   $1.34\pm 0.11$ &   $2.13\pm 0.07$ & 802.74 & $0.34\pm 0.01$ & $2.50\pm 0.05$\\ 

15-day   &   59907.49  & $0.83\pm 0.05$ &    $2.08\pm 0.05$ & 1355.88 & $0.63\pm 0.02$ & $2.64\pm 0.06$  \\

1-month  &   59854.99  &  $0.63\pm 0.03$ &  $1.95\pm 0.04$  & 2792.13 & $0.68\pm 0.02$ & $3.22\pm 0.06$ \\

3-month  &   59884.99  & $0.53\pm 0.02$  &  $2.07\pm 0.02$  & 5729.90 & $0.65\pm 0.02$ & $3.36\pm 0.07$\\

6-month  &   59839.99  &  $0.52\pm 0.01$ &  $2.05\pm 0.02$  & 10315.86& $0.68\pm 0.02$ & $3.43\pm 0.08$ \\

\hline
\end{tabular}
\end{table*}

\begin{figure*}
    \centering
    \includegraphics[scale=0.4,angle=0]{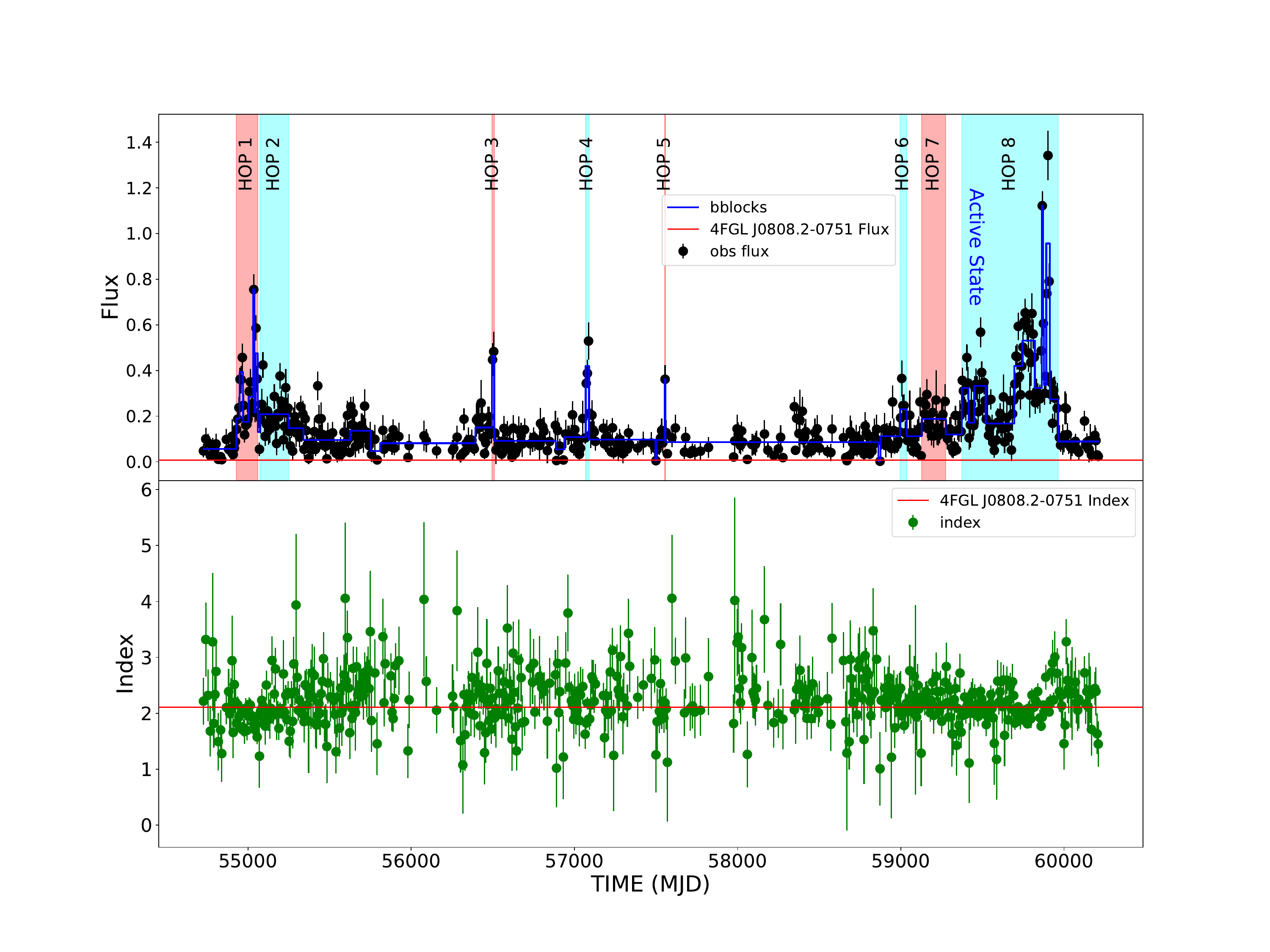}
     \caption {The upper panel depicts the 7-day binned $\gamma$-ray light curve of PKS 0805-07, integrated over the energy range of 0.1–100 GeV [Flux (E$>$100 MeV)] in units of  $~10^{-6}\,\text{photons\, cm}^{-2} \,\text{s}^{-1}$ from MJD 54684 to 60264. The lower panel showcases the corresponding values of  spectral index for the source during the same period. The red horizontal lines in the upper and lower panel represent the flux and index reported in the 4FGL catalog, respectively. The shaded regions represents the HOPs and HOP 8 is designated as ``active state". The different colors representing HOPs are solely for demarcation purposes.} 
     \label{fig:gamma_lc}
\end{figure*}

\begin{figure*}
    \centering
    \includegraphics[scale=0.45,angle=0]{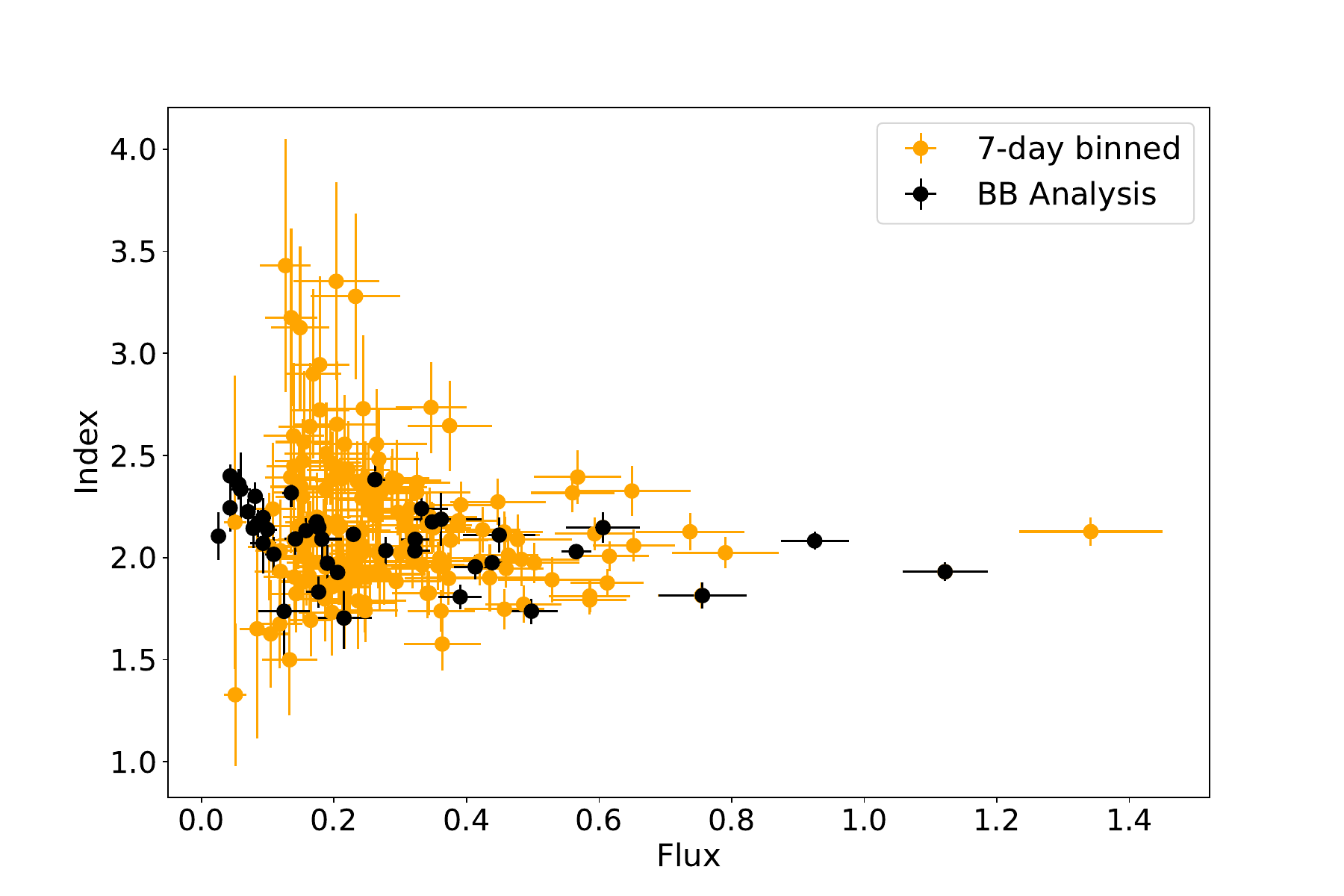}
    \caption{Correlation plot between index and flux in 7-day binned light curve. The yellow solid circles represent the individual flux-index pairs and the black solid circles represent the  flux-index pairs based on BB analysis. Flux is measured in units of in units of  $~10^{-6}\,\text{photons\, cm}^{-2} \,\text{s}^{-1}$.}
    \label{fig:index_flux2}
\end{figure*}

To identify flaring episodes, we followed the approach outlined in \cite{2019ApJ...877...39M}. In this approach, the Bayesian Block  method \citep[BB,][]{2013ApJ...764..167S} is combined with the HOP algorithm \citep[HOP,][]{1998ApJ...498..137E} to  segment a light curve into groups that distinguish between quiescent and flaring episodes. The HOP group is the segment of light curve where the multiple consecutive BBs are connected. In Figure \ref{fig:gamma_lc}, we have shown the variation of 7-day binned flux values and the corresponding index values with respect to the time. We have designated HOP 8 (MJD 59370\textendash{}59965) as the ``active state" of the source. The average flux in this time period is higher than the average flux between MJD 54684\textendash{}59370 (see Table \ref{table:peak} ). Figure \ref{fig:gamma_lc} suggests a negative correlation between the flux and index. To validate this observation, we plotted the spectral index against the 7-day binned $\gamma$-ray flux, applying the condition that the flux/flux-error$>$3. The yellow solid circles in Figure \ref{fig:index_flux2} represent the individual flux-index pairs. Since the lower flux values are associated with larger index errors, we refined the analysis by creating a gamma-ray light curve with time bins corresponding to the BBs derived from the 7-day binned light curve. The flux and index values obtained for each block are also plotted in Figure \ref{fig:index_flux2}, represented by black solid circles. The results indicate that the source demonstrates a mild ``harder when brighter" trend, a common phenomenon observed in blazars \citep{2021MNRAS.508.5921H, 2019MNRAS.484.3168S}. Applying the Spearman rank correlation method, we obtained a correlation coefficient of $-0.25$ with a null-hypothesis probability of $2.62 \times 10^{-5}$ for individual flux-index pairs, and a correlation coefficient of $-0.50$ with a null-hypothesis probability of $6.80 \times 10^{-4}$ for the BB analysis. These results further indicate the mild anti-correlation between the index and flux values.  However, in previous studies e.g., \citet{2021MNRAS.508.5921H, 2019MNRAS.484.3168S},  a clear harder-when-brighter behavior is observed during specific flaring periods.  It is important to note here that the full 7-day binned dataset  includes various flux states of the source, ranging from low to high. Therefore, when analyzing the entire data sample, 
different physical mechanisms may dominate at different flux levels, potentially masking a more obvious correlation. Additionally, the uncertainties related to low-flux states are not considered in the Spearman correlation, so the obtained results should be interpreted with caution.
 Further, the flux and index light curves (see Figure \ref{fig:gamma_lc}) shows a possible lag between the index and flux. Therefore to confirm the lag, we  utilized the Z-transformed discrete correlation function \citep[ZDCF,][]{1997ASSL..218..163A} to find the  possible lag  between flux and index in the time interval MJD 59582--60112 using 3-day binned $\gamma$-ray light curve. We found the index lagging behind the flux by 121 days (+27.2, -3.51). The uncertainties are at 1 $\sigma$ confidence level.

The $\gamma$-ray light curve displays notable variations with number of  low and high flaring components. To undertake a detailed temporal analysis of the source, we chose the 'active state' of the source (see Figure \ref{fig:gamma_lc}). 
 During this time, both the 3-day and 7-day binned $\gamma$-ray light curves exhibit a peak in integral flux. We determined the rise and decay times for these components in this  'active state' using the sum of exponential (SOE) function \citep{2010ApJ...722..520A}:

\begin{equation}\label{eq:rise_fall}
F(t)=F_b+\Sigma F_{i}(t),
\end{equation}

where $F_b$ represents the baseline flux, and

\begin{equation}
F_{i}(t)=\frac{2F_{p,i}}{\exp\left(\frac{t_{p,i}-t}{\tau_{r,i}}\right)+\exp\left(\frac{t-t_{p,i}}{\tau_{d,i}}\right)}\,\,\>\> ,
\end{equation}

\begin{figure*}
    \centering
    \includegraphics[scale=0.45,angle=0]{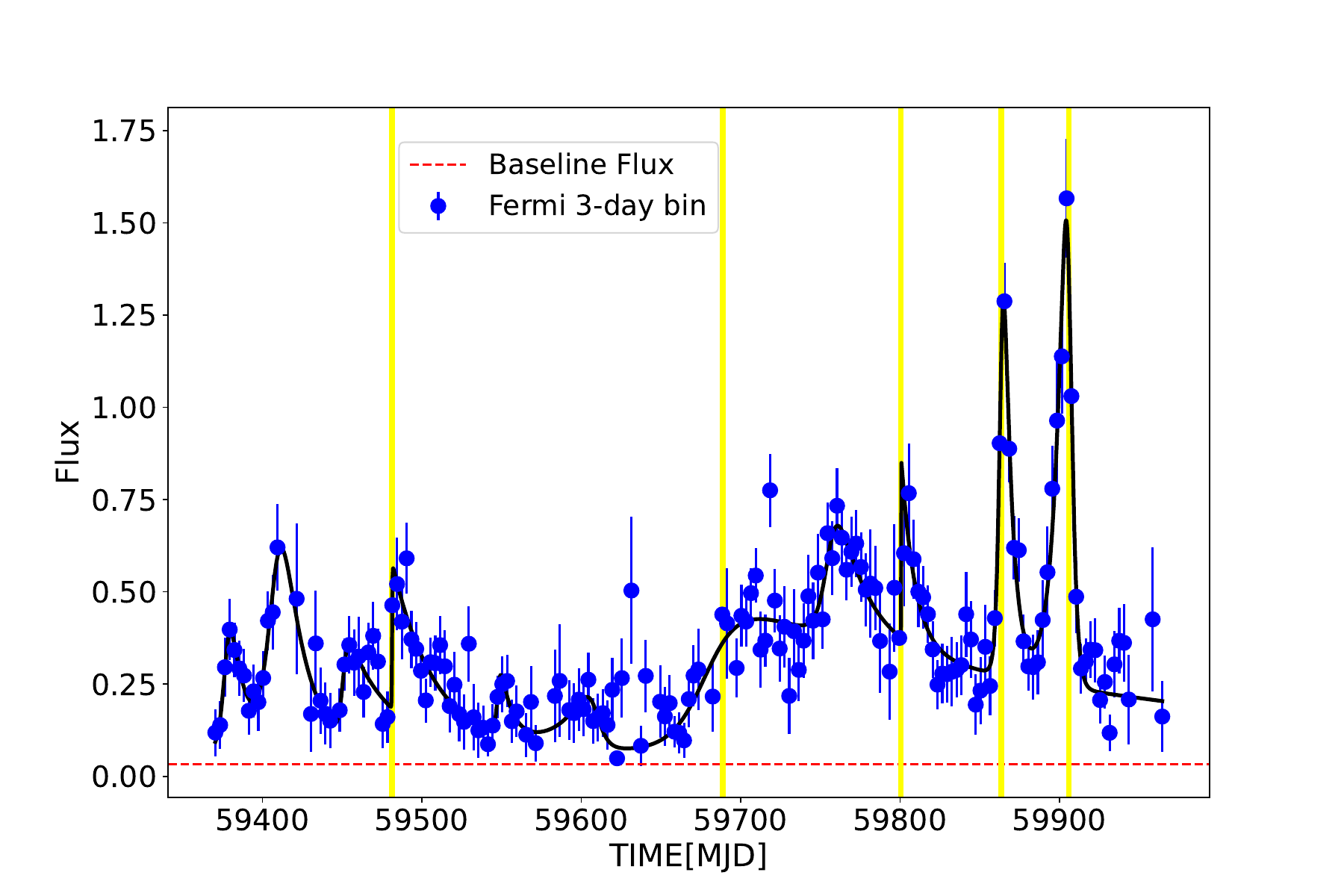}
    \caption{3-day binned $\gamma$-ray light curve of PKS 0805-07 fitted with the SOE function defined in Equation \ref{eq:rise_fall}. Yellow strips denotes the peak time (MJD) of the significant peaks as mentioned in Table \ref{table:multi-comp}. The horizontal red line denotes the baseline flux. Flux is measured in units of in units of  $~10^{-6}\,\text{photons\, cm}^{-2} \,\text{s}^{-1}$.}
    \label{fig:3d_lc_so}
\end{figure*}

In this equation, $F_{p,i}$ is the peak flare amplitude at time $t_{p,i}$, while $\tau_{r,i}$ and $\tau_{d,i}$ denote the rise and decay times of the respective flare component. The fitted SOE profile along with the 3-day binned $\gamma$-ray light curve, are shown in Figure \ref{fig:3d_lc_so}.

We used 11 exponentials in the SOE function, yielding a $\chi^2$/dof value of 123.64/129. Table \ref{table:multi-comp} provides the best fit parameters for components  for which  the ratio of parameter value and its error$>$2. Using rise/decay time scales, we determined the profile shapes of these components by calculating the parameter, $\zeta=\frac{\tau_d-\tau_r}{\tau_d+\tau_r}$, such that the component is symmetric if $|\zeta|<0.3$, moderately asymmetric if $0.3 <|\zeta| < 0.7$, and asymmetric if $0.7 < |\zeta| < 1$ \citep{2010ApJ...722..520A}. We noted that two components are moderately asymmetric and  three components are asymmetric.

\begin{table*}
\caption{The rise and fall  time of the prevailing components in the light curve. Col. 1:  denotes the peak time (MJD), 2:  represents the peak flux (measured in units of $ 10^{-7}\, \text{photons\, cm}^{-2} \,\text{s}^{-1}$), 3 and 4:  display the rise time and decay time of the components (measured in days), and 5: indicates the asymmetry parameter.}
\label{table:multi-comp}
\centering
\begin{tabular}{lcccr}
$ t_p$ & $ F_p$ & $ t_{r}$ & $ t_{d}$ & $|\zeta|$  \\
\hline

$59479.66\pm 2.20\times 10^{-8}$    &   $1.92\pm 0.42$   & $0.05 \pm 9.6\times 10^{-7}$ &   $29.20 \pm 11.17$ & 0.99 \\

$59688.90 \pm 5.98$   &   $ 2.13 \pm 0.30$  &   $11.09 \pm 4.25$ &    $250.48 \pm 56.62$ & 0.92 \\ 

$59800.85 \pm 1.21 \times 10^{-9}$  &   $ 2.45\pm 0.88$  &   $0.02\pm 1.35\times 10^{-14}$ &    8.3$\pm$3.4 & 0.99 \\
$59863.57 \pm 0.77$   &   $8.83 \pm 1.09$   & $1.57 \pm 0.52$ &    $ 4.95 \pm 1.00$   & 0.52  \\

$59905.85 \pm 1.05$  &   $11.31 \pm 1.44$   & $ 6.11 \pm 1.21$ &  $2.23 \pm 0.56$  & 0.46  \\

\hline
  
\end{tabular}
\end{table*}      
        
To estimate the shortest timespan for the doubling of flux, we examined the 3-day binned $\gamma$-ray light curve using the equation:

\begin{equation}
F(t) = F(t_0) \times 2^{\frac{t - t_0}{\tau}},
\end{equation}

here, $F(t_0)$ and $F(t)$ represent the flux values at times $t_0$ and $t$ respectively, and $\tau$ denotes the characteristic doubling time scale. By using the condition that the difference in flux at times  t  and  $t_0$ is significant at a level of ($3\sigma$) or greater \citep{2011A&A...530A..77F},  we identified the shortest time variability as ($2.80 \pm 0.77$) days.

\subsection{Flux Distribution}
The examination of flux distribution of astrophysical systems serves as a valuable tool for investigating the underlying physical processes contributing to variability. For instance, the presence of a normal flux distribution suggests additive processes, while a lognormal distribution indicates multiplicative processes. Typically, the observed flux distribution in compact black hole systems follows a lognormal distribution. To explore this, we characterize the $\gamma$-ray flux distribution of the source by conducting  Anderson-Darling (AD) and histogram fitting tests. The AD test produces a test statistic (TS) value of 13.30, which is well above the critical value (CV) of 0.76 at the 5\%
significance level. Thus AD test rejects the normality of the flux distribution. However, the AD test supports the lognormality of the flux distribution, with a statistic value of 0.54 for the log of the flux distribution, which is smaller than the $CV$ of $0.55$ at the 15\% significance level, suggesting that the null hypothesis of log-normality cannot be rejected. We additionally examined the probability density function (PDF) of the flux distribution by constructing a normalized histogram of the logarithm of flux.

\begin{figure*}
    \begin{center}
        \includegraphics[scale=0.34]{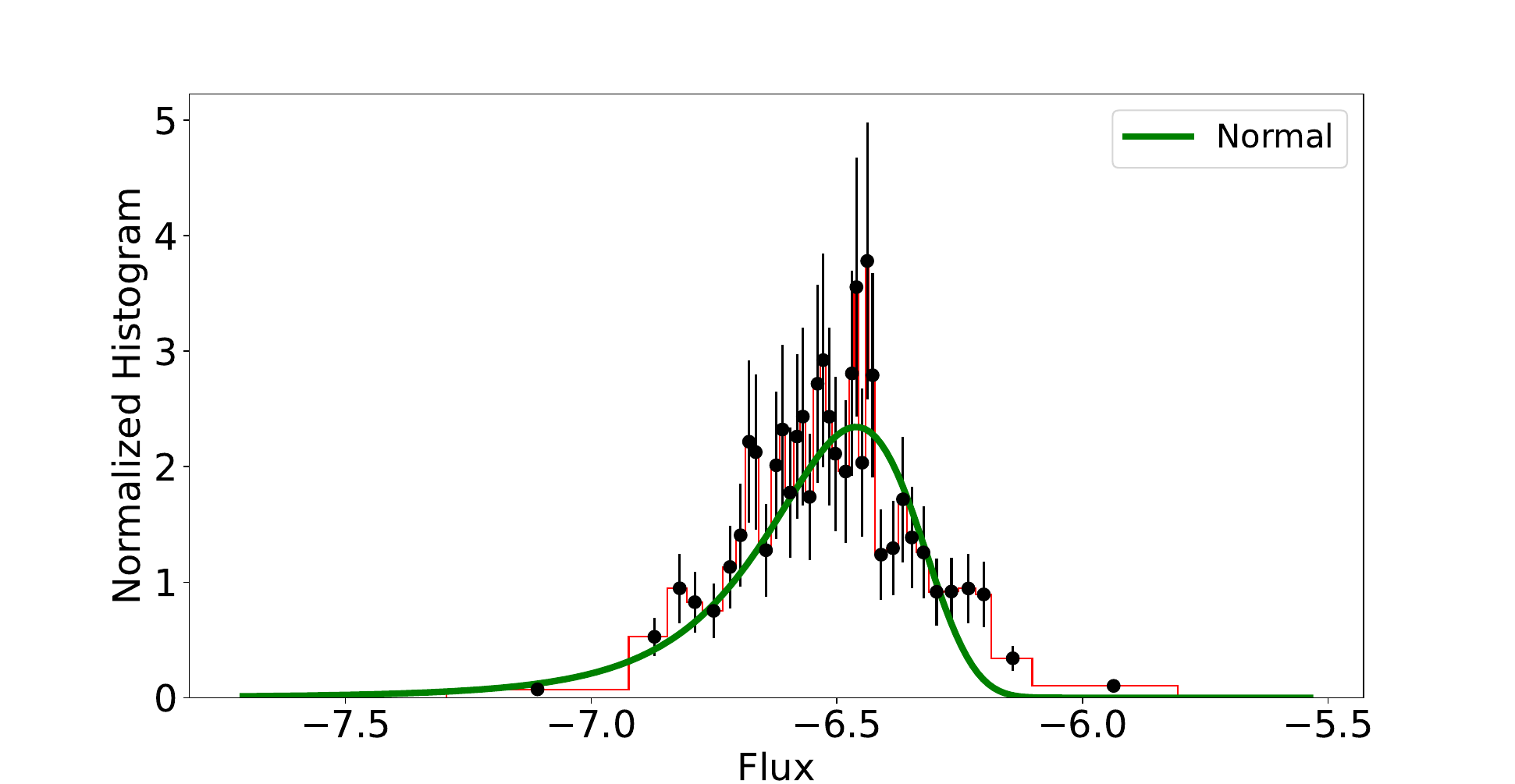}
        \includegraphics[scale=0.34]{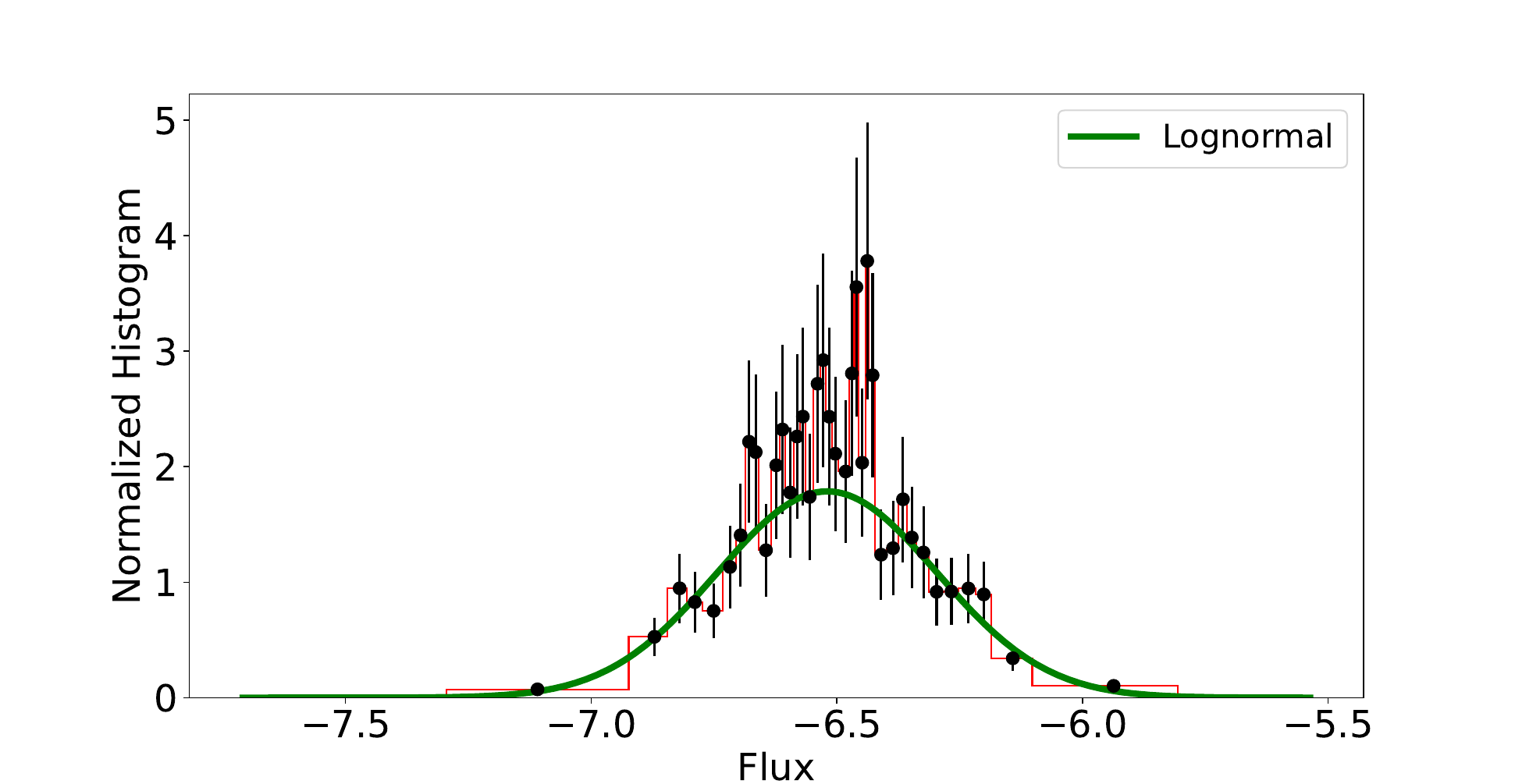}
        \caption{ Normalized histogram of the $\gamma$-ray light curve. Upper panel: normalized histogram fitted using the Gaussian function. Lower panel:  normalized histogram fitted using the lognormal function. Flux is measured in units of  $~10^{-6}\,\text{photons\, cm}^{-2} \,\text{s}^{-1}$}
        \label{hist_fit}
    \end{center}
\end{figure*}

The histogram is constructed with equal points per bin and varying bin widths. For the 3-day binned light curve, there are 370 significant flux points with flux/flux-error$>$3, and each bin contains 10 points. The normalized histogram points are plotted in Figure \ref{hist_fit}. The resulting histogram in log-scale is fitted by:

\begin{equation}\label{eq:ln}
L(x) = \frac{1}{\sqrt{2\pi}\sigma_l}e^{-(x-\mu_l)^2/2\sigma_l^2}
\end{equation}

and

\begin{equation}\label{eq:nor}
G(x) = \frac{10^x \log(10)}{\sqrt{2\pi}\sigma_g}e^{-(10^x-\mu_g)^2/2\sigma_g^2}
\end{equation}

where $\mu_l$ and $\sigma_l$ are the mean and standard deviation of the logarithmic flux distribution, and $\mu_g$ and $\sigma_g$ are the mean and standard deviation of the flux distribution. Equation \ref{eq:ln} represents a lognormal fit, while Equation \ref{eq:nor} represents a normal fit. The lower and upper panels of Figure \ref{hist_fit} show the normalized histogram and the best lognormal/normal fit, respectively. The reduced $\chi^2$ values obtained from fitting the flux distribution with normal and lognormal PDFs were 1.76 and 0.78, respectively, suggesting that the flux distribution is more accurately described by a lognormal distribution. Observation of a lognormal distribution in the $\gamma$-ray light curve implies that the underlying physical processes are likely multiplicative in nature.

\begin{figure*}[!ht]
    \centering
    \includegraphics[width=\textwidth]{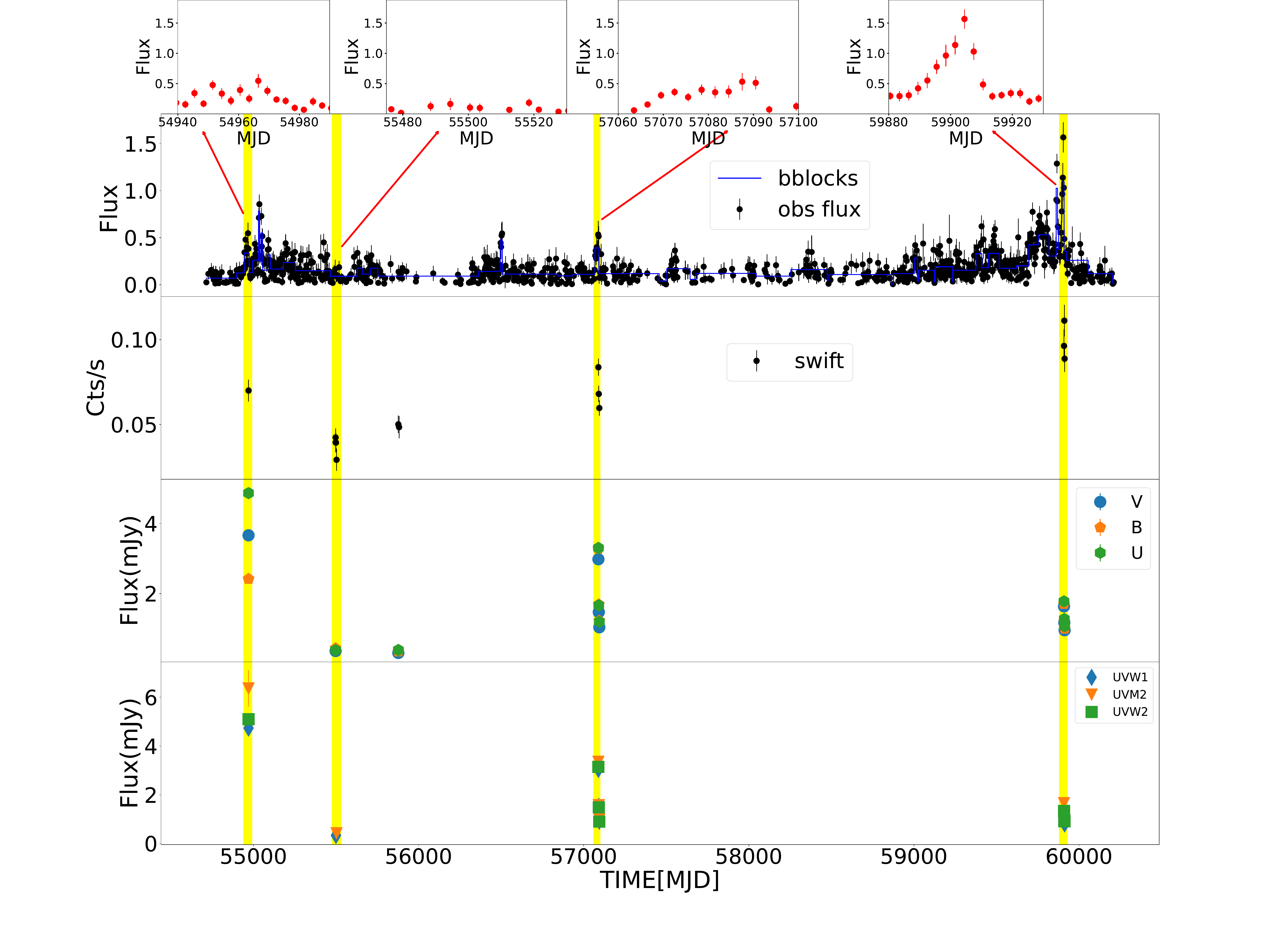}  % Full page width
    \caption{Multiwavelength light curves of PKS 0805-07 obtained using \emph{Fermi}-LAT and \emph{Swift} XRT and UVOT observations. The observations spanned a period from MJD 54684 to 60264. The top panel represents the 3-day binned $\gamma$-ray light curve (Flux is measured in units of  $~10^{-6}\,\text{photons\, cm}^{-2} \,\text{s}^{-1}$ ), and the second, third, and fourth panels depict the X-ray, UV, and optical light curves. Colored vertical stripes indicate the regions where broadband spectral modeling is performed.  The $\gamma$-ray light curve with simultaneous X-ray and optical/UV data is zoomed in and is shown at the top of plot, with arrows indicating their location in the multiwavelength plot.}
    \label{fig:multiplot_lc}
\end{figure*}

\subsection{Multiwavelength Light Curve}

\begin{figure*}
    \begin{center}
        \includegraphics[scale=0.54]{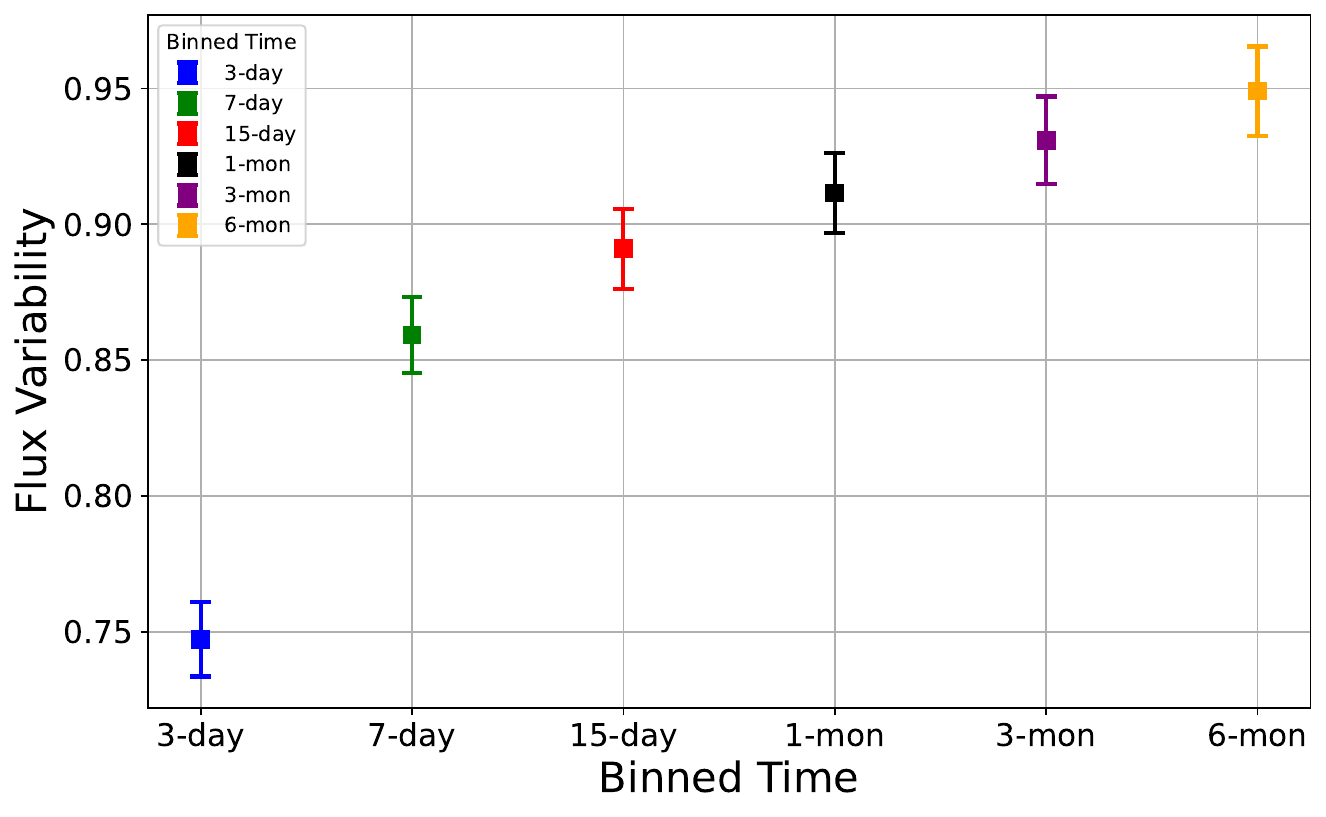}
        
        \caption{Variation of $F_{var}$ with different time binning of $\gamma$-ray light curve.}
        \label{fig:frac_var}
    \end{center}
\end{figure*}

To comprehend the behavior of PKS 0805-07 across the optical, UV, and X-ray bands, all the observations conducted by \emph{Swift}-XRT/UVOT during the period MJD (54684 -- 60264) were taken. The corresponding X-ray and optical/UV data from these observations were analyzed following the procedures outlined in Section \ref{sec:data_ana}. The resulting X-ray and optical/UV light curves for PKS 0805-07 are displayed in the second and bottom panels, respectively, of the multiwavelength light curve plot (refer to Figure \ref{fig:multiplot_lc}). Each data point in the X-ray and optical/UV light curves corresponds to an individual observation. The top panel illustrates the 3-day binned $\gamma$-ray light curve, with flux points derived by integrating over the energy range 0.1--100 GeV.  The part of the $\gamma$-ray light curve with simultaneous X-ray and optical/UV data is zoomed in and is shown at the top of the Figure \ref{fig:multiplot_lc}. The multiwavelength plot shows fluctuations in flux across various energy ranges, but since the data points in X-ray and optical/UV are few, we calculated variability only in $\gamma$-ray light curve by computing the fractional variability amplitude using the formula \citep{2003MNRAS.345.1271V}:

\begin{equation}\label{eq:fvar}
F_{\text{var}} = \sqrt{\frac{S^2 - \overline{\sigma_{\text{err}}^2}}{\overline{F}^2}},
\end{equation}

where $S^2$ represents the variance, $\overline{F}$ is the mean of flux points in the light curve, and $\overline{\sigma_{\text{err}}^2}$ is the mean of the square of the measurement errors. The uncertainty on $F_{\text{var}}$ is determined using the equation \citep{2003MNRAS.345.1271V}:

\begin{equation}
F_{\text{var,err}} = \sqrt{\frac{1}{2N}\left(\frac{\overline{\sigma_{\text{err}}^2}}{F_{\text{var}}\overline{F}^2}\right)^2 + \frac{1}{N}\frac{\overline{\sigma_{\text{err}}^2}}{\overline{F}^2}},
\end{equation}

where $N$ is the number of points in the light curve.  The $F_{var}$ values for the $\gamma$-ray light curve across different time bins are shown in Figure \ref{fig:frac_var}, where we observed an increase in $F_{var}$ with larger bin sizes. In our analysis, we applied a $TS > 4$ threshold to include only statistically significant data, reducing the risk of bias from less reliable flux points.  \citet{2019Galax...7...62S} reported an opposite trend for Mrk\,421, which result from including all flux points, regardless of their statistical significance. This suggests that $F_{var}$ is highly sensitive to the data selection criteria. 
%We consider the $TS > 4$ threshold, as it filters out potentially inaccurate data and ensures that our results are based on statistically significant flux measurements.

\subsection{Broadband Spectral Analysis}
To understand the  physical parameters responsible for simultaneous flux variations, we examine the broadband spectral characteristics of  PKS 0805-07 by dividing the $\gamma$-ray light curve into segments using BB analysis. These segments, shown in Figure \ref{fig:multiplot_lc}, 
represent distinct flux states (active and quiescent states).  In our work, we selected the states for which the simultaneous observations are available in the X-ray and Optical/UV bands. We have illustrated these temporal intervals using vertical stripes in the multiplot (Figure \ref{fig:multiplot_lc}) and categorized them as follows: S1 MJD\,54940-54990, Q1 MJD\,55475-55530, S2 MJD\,57060-57100, and S3 MJD\,59880-59930. S1, S2 and S3 correspond to active states, while Q1 represents a quiescent state.  We characterized the $\gamma$-ray spectrum of the selected flux states using the log parabola (LP) model and powerlaw (PL) model. The LP and PL models are expressed as follows:
\begin{equation}
    \frac{dN}{dE} = N_0 \frac{E}{E_0}^{-(\alpha+\beta \log(E/E_0))} \, ,
\end{equation}

where $N_0$ represents the normalization, $\alpha$ is the photon index at the scale energy, $E_0$ is fixed at 665.74 MeV, and the parameter $\beta$ measures the curvature in the spectrum.

\begin{equation}
    \frac{dN}{dE} = N_0 \left(\frac{E}{E_0}\right)^{-\Gamma}
\end{equation}

Here, $ N_0$ is the prefactor, $\Gamma$ denotes the spectral index, and $E_0$ is the fixed scale energy set at 733 MeV. The obtained fitting parameters are summarized in Table \ref{table:spec_fit_param}.
We evaluated the statistical significance of the observed curvature in the $\gamma$-ray spectrum using the relation $TS_{curve}=2[\log\mathcal{L}(LP)-\log\mathcal{L}(PL)]$ \citep{2012ApJS..199...31N}. As mentioned in Table \ref{table:spec_fit_param}, significant curvature ($TS_{curve}>16$) is observed in the S2 and S3 states.

\begin{table*}
\centering
    \caption{The parameters obtained by fitting the integrated $\gamma$-ray spectrum of S1, Q1, S2, and S3 states of PKS\,0805-07 with the PL and LP model. Col. 1: flux state; 2: time period of flux state; 3: fitted model; 4: integrated flux in units of $~10^{-7}\,\text{photons\, cm}^{-2} \,\text{s}^{-1}$; 5: PL index or index defined at pivot energy; 6: curvature parameter; 7: test statistics; 8: -log(likelihood); 9: significance of curvature.}
    \vspace{0.5cm}  % Adds some space between the caption and the table
    \begin{tabular}{lcccccccr}
        \toprule
        State & Period & Model & $\rm F_{0.1-300}$ & $\Gamma$ or $\alpha$ & $\beta$ & TS & -$\log\mathcal{L}$ & $\rm TS_{curve}$ \\
        \midrule
        S1 & 54940-54990 & PL & $2.91\pm 0.18$ & $1.90\pm 0.04$  & ---  & 1748  &  17566 & ---  \\
         &  & LP & $2.60\pm 0.21$ & $1.8\pm0.06$ & $0.06\pm0.02$ & 1775 & 17563 &  6 \\
        Q1 & 55475-55530 & PL & $1.07\pm0.12$ & $2.17\pm0.11$ & --- & 125 & 13966  & --- \\
        & & LP & $0.74\pm 0.19$  & $2.09\pm 0.16$ & $0.08\pm0.08$ & 130 & 13966 & 0 \\
        S2 & 57060-57100 & PL & $3.44\pm0.23$ & $2.04\pm0.04$ & --- & 966 & 12476 & --- \\
        &  & LP  & $2.81\pm0.26$ & $1.89\pm0.07$ & $0.17\pm0.04$ & 1040 & 12464  &  24 \\
        S3 & 59880-59930 & PL & $5.61\pm0.26$ & $2.21\pm0.03$ & --- & 1813 & 18370 & --- \\
        &  & LP  & $5.4\pm0.0.03$ & $2.17\pm0.04$ & $0.14\pm0.03$ & 1855 & 18357 & 26 \\ 
        \bottomrule
    \end{tabular}
\label{table:spec_fit_param}
\end{table*}

We performed broadband spectral analysis on these segments to identify changes in physical parameters governing flux variations. We adopt a one-zone leptonic model to characterize the broadband SED during selected flux states. In this model, we consider emission originating from a spherical blob with radius $\emph{R}$ filled with a relativistic electron distribution, $ n(\gamma)$. The blob moves along the jet with a bulk Lorentz factor $\Gamma$ at a small angle $\theta$ relative to the observer's line of sight. The beaming factor, $\delta = 1/\Gamma(1-\beta\cos\theta)$, amplifies the blazar emission due to relativistic motion. We assume variability is governed by light crossing time scales, determining the emission region size as $ R\sim \delta t_{var}/(1+z)$. The relativistic electrons, in the presence of magnetic field $\rm B$ and target photon field, emit radiation through synchrotron and Inverse Compton (IC) processes. We assume seed photons for the IC process are synchrotron photons from the jet itself, resulting in emission through Synchrotron Self-Compton (SSC) process. We express the electron Lorentz factor $\gamma$ in terms of a new variable $\xi$ such that $\xi  =\gamma\sqrt{\mathbb{C}}$, where $ \mathbb{C} =1.36\times 10^{-11}\delta B/(1+z)$. Following \citet{1984RvMP...56..255B, 1995ApJ...446L..63D,  2008ApJ...686..181F, 2018RAA....18...35S}, the synchrotron flux at energy $\epsilon$ can be obtained using the equation

\begin{equation}\label{eq:syn_flux}
F_{\mathrm{syn}}(\epsilon)=\frac{\delta^3(1+z)}{d_L^2} V \mathbb{A} \int_{\xi_{\mathrm{min}}}^{\xi_{\mathrm{max}}} f(\epsilon/\xi^2)n(\xi)d\xi,
\end{equation}

where $\mathrm{d_L}$ represents the luminosity distance, $V$ is the emission region volume , $\mathbb{A} = \frac{\sqrt{3}\pi e^3 B}{16m_e c^2 \sqrt{\mathbb{C}}}$. The symbols $\xi_{\mathrm{min}}$ and $\xi_{\mathrm{max}}$ represenat the minimum and maximum energy of electrons, respectively. Additionally, the function $f(x)$ denotes the synchrotron emissivity function, as described in \citep{1986rpa..book.....R}.

The observed SSC flux at energy $\epsilon$ can be calculated using the following expression \citet{2008ApJ...686..181F, 2018RAA....18...35S}:

\begin{equation}\label{eq:ssc_flux}
\begin{split}
F_{\mathrm{ssc}}(\epsilon) =\frac{\delta^3(1+z)}{d_L^2} V \mathbb{B} \epsilon & \int_{\xi_{\mathrm{min}}}^{\xi_{\mathrm{max}}} \frac{1}{\xi^2} \int_{x_1}^{x_2} \frac{I_{\mathrm{syn}}(\epsilon_i)}{\epsilon_i^2} \\
& f(\epsilon_i, \epsilon, \xi/\sqrt{\mathbb{C}}) d\epsilon_i n(\xi)d\xi
\end{split}
\end{equation}

where  $\epsilon_i$ corresponds to incident photon energy  , $\mathbb{B} = \frac{3}{4}\sigma_T\sqrt{\mathbb{C}}$, $I_{\mathrm{syn}}(\epsilon_i)$ represents the synchrotron intensity, $x_1=\frac{\mathbb{C} \, \epsilon}{4\xi^2(1-\sqrt{\mathbb{C}} \,\epsilon/\xi m_ec^2)}$, $x_2=\frac{\epsilon}{(1-\sqrt{\mathbb{C}}\,\epsilon/\xi m_e c^2)}$, and

\begin{equation}
f(\epsilon_i, \epsilon, \xi)= 2q\log q+ (1+2q)(1-q)+\frac{\kappa^2q^2(1-q)}{2(1+\kappa q)} \nonumber
\end{equation}

where $q=\frac{\mathbb{C}\epsilon}{4\xi^2\epsilon_i(1-\sqrt{\mathbb{C}}\epsilon/\xi m_ec^2)}$ and $\kappa=\frac{4\xi\epsilon_i}{\sqrt{\mathbb{C}} m_e c^2}$.

In a similar manner, the observer can determine the observed EC flux  through the following equation \citep{2018RAA....18...35S}:

\begin{equation}\label{eq:ec_flux}
\begin{split}
F_{\mathrm{ec}}(\epsilon) =\frac{\delta^3(1+z)}{d_L^2} V \mathbb{D} \epsilon
 \int_{0}^{\infty} d\epsilon_i^* \int_{\xi_{\mathrm{min}}}^{\xi_{\mathrm{max}}} d\xi \frac{N(\xi)}{\xi^2} \frac{U_{\mathrm{ph}}^*}{\epsilon_i^*} \eta(\xi,\epsilon_s, \epsilon_i') \\
\end{split}
\end{equation}

Where $\mathbb{D}=\frac{3}{32\pi}c\beta\sigma_T\sqrt{\mathbb{C}}$, $\epsilon_i^*$ represents the energy of target photons in the AGN frame, $U_{ph}^*$ signifies the energy density of target photons, 
\begin{equation}
\eta(\xi,\epsilon_s, \epsilon_i') =y+\frac{1}{y}+\frac{\mathbb{C}\epsilon_s^2}{\xi^2\epsilon_i'^2y^2}-\frac{2\nu_s\sqrt{\mathbb {C}}}{\xi\epsilon_i'y}
\end{equation}
where $y=1-\frac{\sqrt{\mathbb{C}} \epsilon_s}{\xi m_e c^2}$.

\begin{figure*}
		\begin{center}
	\includegraphics[angle=270,width=.47\textwidth]{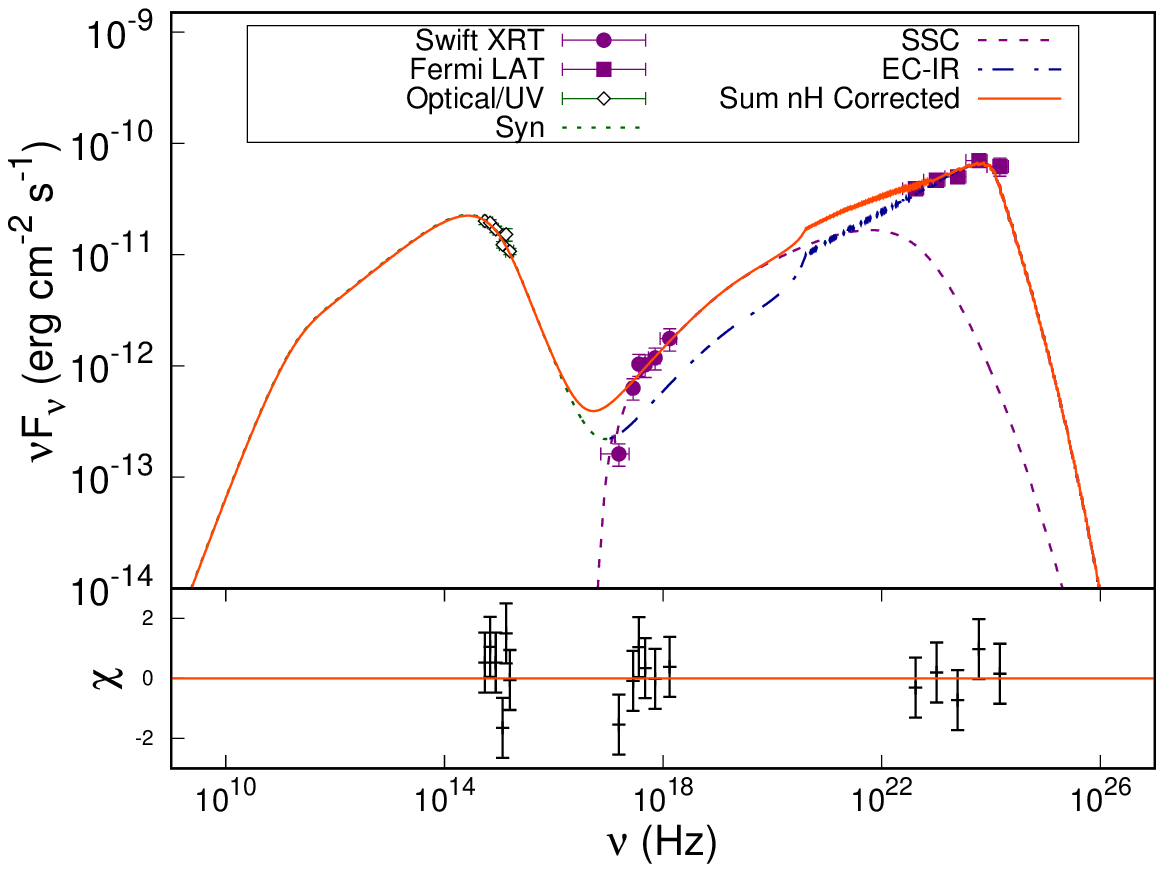}
        \includegraphics[angle=270,width=.47\textwidth]{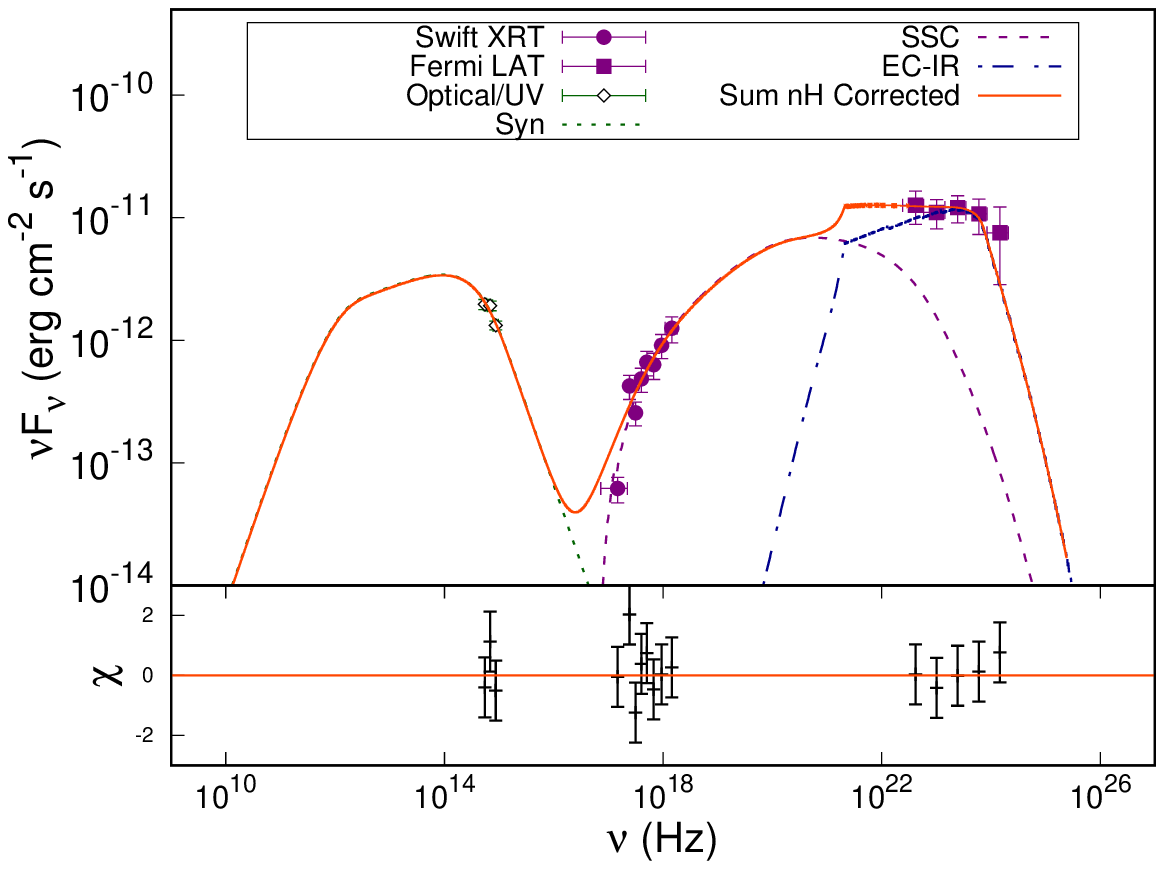}
        \includegraphics[angle=270,width=.47\textwidth]{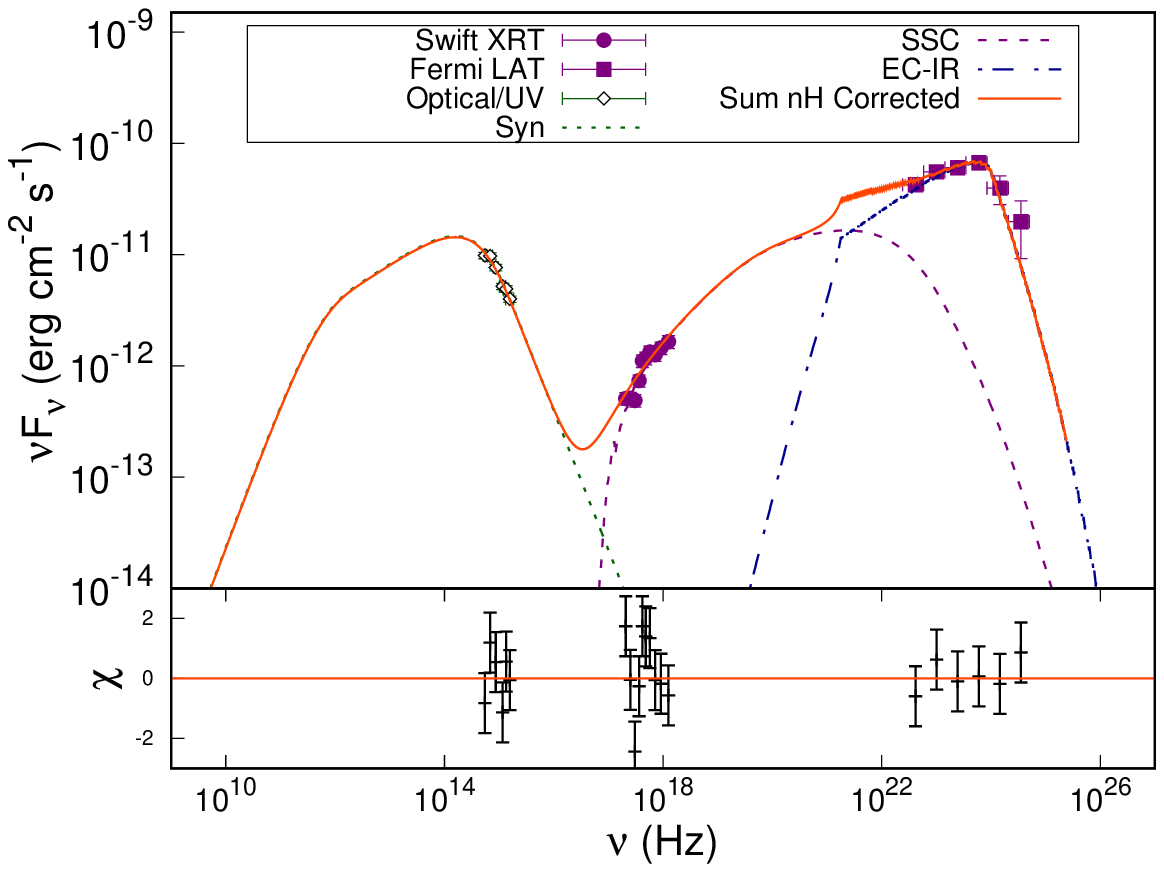}
        \includegraphics[angle=270,width=.47\textwidth]{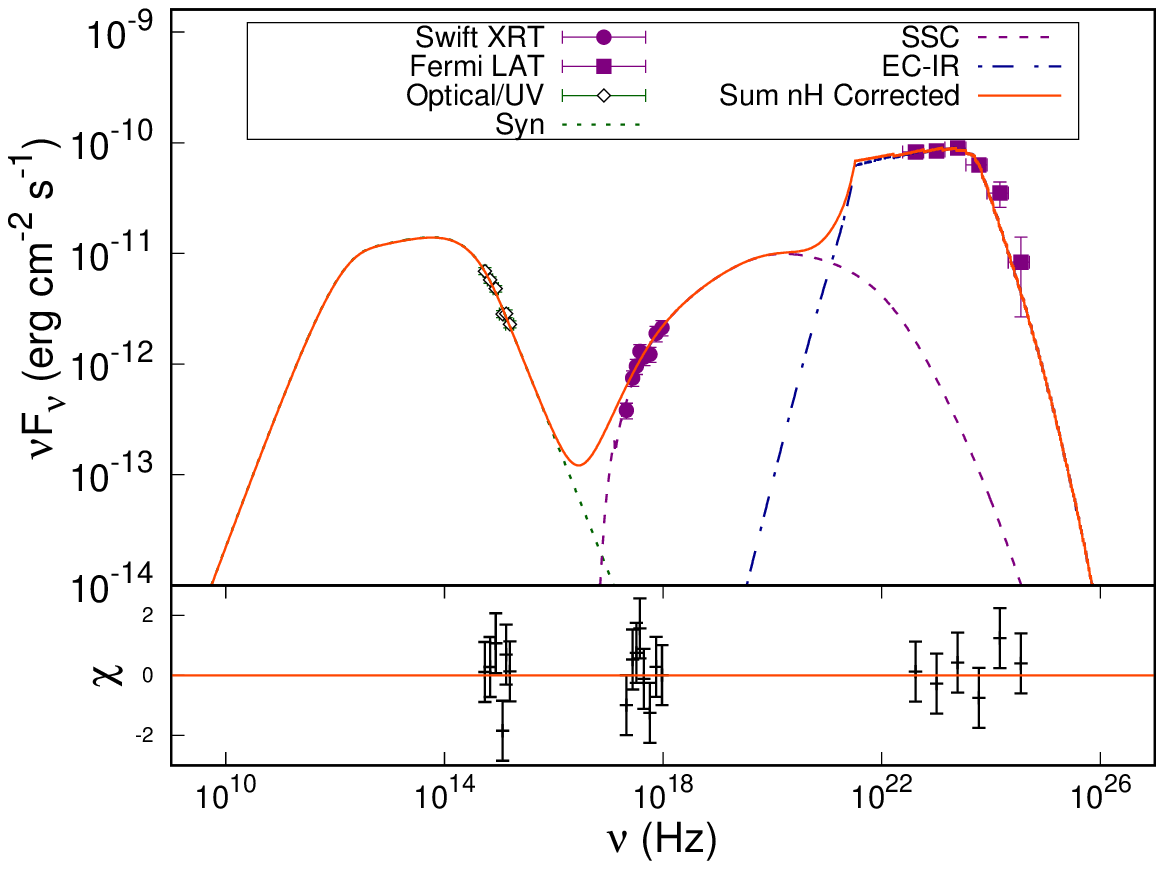}
	   \vspace{0.9cm}
        \caption{The broadband SED of PKS\,0805-07 acquired during the flux states S1 (on the top left panel), Q1 (on the top right panel), S2 (on the bottom left panel) and S3 (on the bottom right panel). The flux points denoted by filled diamonds correspond to \emph{Swift}-UVOT, squares correspond to \emph{Swift}-XRT, and filled circles correspond to  \emph{Fermi}-LAT. The solid red curve illustrates the combined best-fit  synchrotron, SSC, and EC spectra.}

        \label{fig:s12}
		\end{center}        
\end{figure*}

We solved Equations \ref{eq:syn_flux}, \ref{eq:ssc_flux} and \ref{eq:ec_flux} numerically. for a broken power law electron distribution The resulting numerical code is incorporated as a local convolution model in XSPEC for statistical fitting of broadband SEDs. 
 A systematic error of 5\% was added individually to UVOT. The observed broadband spectrum is mainly determined by 10 parameters: $ \xi_b$, $ \xi_{min}$, $ \xi_{max}$, p, q, $\Gamma$, B, R, $\theta$, and norm N. The code also allows fitting the SED with jet power ($P_{jet}$) as one of the parameters, with N fixed. The initial parameter values were selected based on the shape and flux levels of the synchrotron/SSC/EC components during the flaring state. Subsequently, we varied the parameters individually to identify the optimal values. After obtaining the optimal parameter values, we performed the final fit, allowing only four parameters i.e., p, q, $\Gamma$ and B  to vary, while fixing the others, including N, at their optimal values. The reason for freezing certain parameters is the limited information available in the Optical/UV, X-ray, and $\gamma$-ray bands. Moreover, we used the \texttt{Tbabs} model to account for absorption in the X-ray spectrum.
We observed that the synchrotron,  SSC and EC emissions yielded a satisfactory fit for all flux states, resulting in $ \chi^2/dof$ values of 12.43/12, 9.15/12, 21.74/17 and 13.59/15 for S1, Q1, S2 and  S3  states respectively. The resulting SED best-fit model, along with observed points, are depicted in Figure \ref{fig:s12} and the corresponding optimal parameters are provided in Table \ref{tab:Sed_parameters1}.

\begin{table*}
    \centering
    \caption{Broadband SED model parameters of PKS\,0805-07 for various flux states. Col. 1: Low energy particle index, 2: High energy particle index, 3: Magnetic field in units of $10^{-3}$ Gauss, 4: Bulk Lorentz factor of the emission region, 5: $\chi^{2}/dof$, 6: Logarithmic jet power in units of erg $s^{-1}$, 7: Normalization. The size of the emission region ($R$) was fixed at $10^{17}$cm, $\xi_{\max}$ fixed at $10$, viewing angle ($\theta$) at $2^{\circ}$, and target photon temperature at 1000 K. The $\xi_{min}$ parameter ranges between ($0.7-1.9 )\times 10^{-3}$. The fraction of seed photons undergoing IC scattering ranges from 0.2-0.4 percent. The values in subscript and superscript for parameters in the model represent their lower and upper values, respectively, obtained through the broadband spectral fitting.}
    
    \renewcommand{\arraystretch}{2.5}
    \begin{tabular}{lcccccccc}
        \hline
        & \multicolumn{8}{c}{Free Parameters} \\ \cline{2-9}  % Adjusted the cline command to cover all columns
        State & $\xi_{brk}$ & p & q & B & $\Gamma$ & $\chi^{2}/dof$ & $P_{jet}$ & N \\ \hline  % Fixed alignment of headers
        S1 & $4.20\times10^{-2}$ & $2.20^{2.24}_{2.16}$ & $5.49^{6.10}_{5.02}$ & $0.27^{0.28}_{0.25}$ & $22.86^{24.08}_{21.84}$ & 12.43/12 & 46.30 & $9.18\times10^{-10}$ \\
        Q1 & $3.24\times10^{-2}$ & $2.66^{2.82}_{2.49}$ & $5.58^{6.53}_{4.31}$ & $0.18^{0.23}_{0.15}$ & $10.15^{11.84}_{8.29}$ & 9.15/12 & 45.62 & $2.43\times10^{-9}$ \\ 
        S2 & $3.4\times10^{-2}$ & $2.34^{2.39}_{2.30}$ & $5.52^{5.90}_{5.16}$ & $0.20^{0.21}_{0.18}$ & $19.71^{20.22}_{18.92}$ & 21.74/17 & 46.09 & $1.23\times10^{-8}$ \\
        S3 & $2.91\times10^{-2}$ & $2.76^{2.81}_{2.70}$ & $5.47^{5.79}_{5.17}$ & $0.26^{0.29}_{0.25}$ & $21.66^{22.22}_{21.13}$ & 13.59/15 & 46.10 & $1.93\times10^{-9}$ \\ 
        \hline
    \end{tabular}
    \label{tab:Sed_parameters1}
\end{table*}

\section{Summary and Discussion}
\label{sum:dis}
The continuous monitoring of the FSRQ PKS 0805-07 by Fermi-LAT, coupled with simultaneous observations from Swift-XRT/UVOT, has enabled  a comprehensive examination of the source's temporal and spectral characteristics. For the first time, a detailed analysis of the source was carried out in this paper.  We have designated HOP 8 (MJD 59370\textendash{}59965) as the ``active state" of the source. The average flux in this time period is higher than the average flux between MJD 54684\textendash{}59370 (see Table \ref{table:peak}). The maximum  $\gamma$-ray flux in the 3-day and 7-day binned light curves was obtained as  $(1.56\pm 0.16)\times10^{-6}  \text{photons\, cm}^{-2} \,\text{s}^{-1}$ on MJD 59904.5 and $(1.34\pm 0.11)\times10^{-6}\, \text{photons\, cm}^{-2} \,\text{s}^{-1}$  on MJD 59901.5, respectively. During the ``active state", the $\gamma$-ray light curve revealed the presence of multiple flaring components.  The 3-day binned $\gamma$-ray light curve showed a total of 11 flaring components during the active state (see Figure \ref{fig:3d_lc_so}). We noted that while two components are moderately asymmetric and the other  three components are asymmetric (see Table \ref{table:multi-comp}). The asymmetry observed in the flare profile could be attributed to fluctuations in the strength of the acceleration process. A gradual rise in the asymmetric flare potentially indicates the acceleration of particles to higher energy levels, while a rapid decay may be associated with the swift energy loss of high-energy particles.  The 3-day binned $\gamma$-ray light curve revealed the shortest  flux doubling  time scale of $\mathrm {t_{var}=2.80\pm 0.77}$ days.  Further, we utilized the Z-transformed discrete correlation functioFn \citep[ZDCF,][]{1997ASSL..218..163A} to find the correlation and possible lag between flux and index in the time interval MJD 59582 to 60112 using a 3-day binned $\gamma$-ray light curve. A consistent trend of spectral hardening with increased brightness was observed in the $\gamma$-ray light curve. Moreover, we noted that the index lags behind the flux by 121 (+27.2, -3.51) days. 

We observed that  $F_{var}$  increases with larger bin sizes. In our analysis, we applied a $TS > 4$ threshold to include only statistically significant data points. However, it is important to note that excluding data points with $TS<4$ for Fermi light curves, which typically have the same exposure in each observation, effectively removes low-flux points. This may introduces a bias in the $F_{var}$ calculation, as removing these points reduces the variability amplitude. Additionally,  in the lower time-binned light curves (e.g., 3-day or 7-day), the fraction of points having $TS<4$ are larger compared to higher time binned light curves (e.g., 3 or 6-month). Consequently, more data points are excluded from the 3-day or 7-day binned light curves, resulting in a noticeable decrease in the $F_{var}$ values, as shown in Figure \ref{fig:frac_var}. This contrasts with the trend observed when all measurements are included, where $F_{var}$ increases with the decrease in size of  time bin. As such, the inclusion of all data points, despite their uncertainties, reveals an opposite trend in the time-binned variability \citep{2019Galax...7...62S}.

We examined the flux distribution of the $\gamma$-ray light curve using  Anderson-Darling (AD) test  and histogram fitting.  These tests reject the assumption of normality in the flux distribution and instead suggest that the flux distribution follows a lognormal pattern. The observation of a lognormal distribution implies that the underlying emission process responsible for the variability is multiplicative rather than additive \citep[see][and reference theirin]{2018RAA....18..141S}. 
For blazars, where the emission mainly originates from the jet, a potential explanation for the observation of lognormal behavior is that the disk fluctuations are imprinted on the jet emission \citep{2009A&A...503..797G, 2010LNP...794..203M, 2018RAA....18..141S}. Conversely, the minute timescale variations observed in $\gamma$-ray light curves suggest that the jet emission should be independent of accretion disk fluctuations \citep{2012MNRAS.420..604N}. In such cases, the lognormal distribution in flux can be explained by linear Gaussian perturbations in particle acceleration timescales \citep{2018MNRAS.480L.116S}.  Alternatively, a log-normal flux distribution can also arise from additive processes under certain conditions. For instance, if the blazar jet is considered as a collection of mini-jets, the logarithm of the composite flux would exhibit a normal distribution \citep{2012A&A...548A.123B}.
 
The spectral examination of the $\gamma$-ray light curve within the specified flux states reveals significant curvature in the S2 (TS curve = 24) and S3 (TS curve = 26) states. The conventional approach to comprehend spectral curvature involves considering radiative losses in the emission region, as proposed by \citet{2002MNRAS.336..721K}.  
 Alternatively, \citet{2004A&A...422..103M} demonstrated that curved features in particle distributions may arise when the acceleration probability is energy-dependent. Another explanation involves the energy dependency of the escape timescale in the acceleration zone, as proposed by \citet{2018MNRAS.480.2046G,2021MNRAS.508.5921H,2022MNRAS.515.3749K}. Additionaly, \citet{2021MNRAS.508.5921H} demonstrated that models incorporating power-law (PL) with maximum energy, energy-dependent diffusion (EED), and energy-dependent acceleration (EDA) can generate spectral curvature. However, they ruled out the PL with maximum energy model due to inconsistencies between observed correlations and model predictions. 

 In this paper, we employed the one-zone leptonic model to reproduces the broadband spectra of the selected flux states.  We assumed the  steady-state emission within the chosen flux intervals. Our modeling relied on a BPL  electron distribution, extensively utilized for SED modeling in prior research \citep{2018RAA....18...35S, 2024MNRAS.527.5140S}. This BPL electron distribution, subject to synchrotron, SSC, and EC losses, successively reproduces the broad-band emission across all flux intervals. We observed that the EC scattering of IR target photons yielded a satisfactory fit to the data. 
 The statistical fitting process involves treating p, q, $\Gamma$, $\xi_{brk}$ and $B$ as free parameters, while the remaining parameters are held constant at typical values specific to the corresponding flux state. The corresponding optimal parameters are detailed in Table \ref{tab:Sed_parameters1}. It is observed that the magnetic field strength $B$ varies within the range of (0.18 to 0.27)$\times 10^{-3}$ Gauss, showing an upward trend from low flux state to high flux state.  Notably, such low magnetic field values, measured in milligauss, have also been reported in radio observations of FSRQs  \citep[e.g.][]{2022MNRAS.510..815K,2023MNRAS.523.5703J}.
 
 Similarly, the Lorentz factor $\Gamma$ ranges from 10.15-22.86, exhibiting an increase from Q1 to S3, with the highest $\Gamma$ observed in state S1. 
Furthermore, there exists a positive correlation between $B$ and $\Gamma$. The break energy, defined as $\xi_{break}$, ranges between (2.91 to 4.20)$\times 10^{-2}$. Nevertheless, there is no clear  trend observed from the low to high flux states.
 The derived spectral indices, denoted as $p$ and $q$, vary between 2.20 - 2.76 and 5.47 - 5.58, respectively. These index values suggest a steeper spectral slope beyond the break energy than expected solely from synchrotron cooling, necessitating alternative explanations. The precise cause for this steepened spectrum remains ambiguous. One plausible explanation could involve the presence of an energy-dependent diffusion coefficient. Earlier investigations, such as those by \citet{2007A&A...465..695Z}, have shown that with an energy-dependent diffusion coefficient, the spectral cutoff adopts a sub-exponential or steeper profile at higher energies. Alternatively, \citet{2008MNRAS.388L..49S} demonstrated that a two-zone model incorporating a BPL injection into cooling region can lead to a steeper index after the break energy.

%% Also note that the akcnowlodgment environment does not support long amounts of text. If you have a lot of people and institutions to acknowledge, do not use this command. Instead, create a new \section{Acknowledgments}.
\begin{acknowledgments}
We sincerely thank the anonymous referee for his valuable feedback and suggestions, which have significantly improved the quality of our manuscript.   SAD is thankful to the MOMA for the MANF fellowship (No.F.82-27/2019(SA-III)). ZS is supported by the Department of Science and Technology, Govt. of India, under the INSPIRE Faculty grant (DST/INSPIRE/04/2020/002319). SAD, ZS and  NI express  gratitude to the Inter-University Centre for Astronomy and Astrophysics (IUCAA) in Pune, India, for the support and facilities provided.

\end{acknowledgments}


\vspace{5mm}


% \software{
%           }




\bibliography{sample631}{}

\begin{thebibliography}{}
\expandafter\ifx\csname natexlab\endcsname\relax\def\natexlab#1{#1}\fi
\providecommand{\url}[1]{\href{#1}{#1}}
\providecommand{\dodoi}[1]{doi:~\href{http://doi.org/#1}{\nolinkurl{#1}}}
\providecommand{\doeprint}[1]{\href{http://ascl.net/#1}{\nolinkurl{http://ascl.net/#1}}}
\providecommand{\doarXiv}[1]{\href{https://arxiv.org/abs/#1}{\nolinkurl{https://arxiv.org/abs/#1}}}

\bibitem[{{Abdo} {et~al.}(2010{\natexlab{a}}){Abdo}, {Ackermann}, {Ajello},
  {Allafort}, {Antolini}, {Atwood}, {Axelsson}, {Baldini}, {Ballet},
  {Barbiellini}, {Bastieri}, {Baughman}, {Bechtol}, {Bellazzini}, {Belli},
  {Berenji}, {Bisello}, {Blandford}, {Bloom}, {Bonamente}, {Bonnell},
  {Borgland}, {Bouvier}, {Bregeon}, {Brez}, {Brigida}, {Bruel}, {Burnett},
  {Busetto}, {Buson}, {Caliandro}, {Cameron}, {Campana}, {Canadas}, {Caraveo},
  {Carrigan}, {Casandjian}, {Cavazzuti}, {Ceccanti}, {Cecchi}, {{\c{C}}elik},
  {Charles}, {Chekhtman}, {Cheung}, {Chiang}, {Cillis}, {Ciprini}, {Claus},
  {Cohen-Tanugi}, {Conrad}, {Corbet}, {Davis}, {DeKlotz}, {den Hartog},
  {Dermer}, {de Angelis}, {de Luca}, {de Palma}, {Digel}, {Dormody}, {Silva},
  {Drell}, {Dubois}, {Dumora}, {Fabiani}, {Farnier}, {Favuzzi}, {Fegan},
  {Ferrara}, {Focke}, {Fortin}, {Frailis}, {Fukazawa}, {Funk}, {Fusco},
  {Gargano}, {Gasparrini}, {Gehrels}, {Germani}, {Giavitto}, {Giebels},
  {Giglietto}, {Giommi}, {Giordano}, {Giroletti}, {Glanzman}, {Godfrey},
  {Grenier}, {Grondin}, {Grove}, {Guillemot}, {Guiriec}, {Gustafsson},
  {Hadasch}, {Hanabata}, {Harding}, {Hayashida}, {Hays}, {Healey}, {Hill},
  {Horan}, {Hughes}, {Iafrate}, {J{\'o}hannesson}, {Johnson}, {Johnson},
  {Johnson}, {Johnson}, {Kamae}, {Katagiri}, {Kataoka}, {Kawai}, {Kerr},
  {Kn{\"o}dlseder}, {Kocevski}, {Kuss}, {Lande}, {Landriu}, {Latronico}, {Lee},
  {Lemoine-Goumard}, {Lionetto}, {Llena Garde}, {Longo}, {Loparco}, {Lott},
  {Lovellette}, {Lubrano}, {Madejski}, {Makeev}, {Marangelli}, {Marelli},
  {Massaro}, {Mazziotta}, {McConville}, {McEnery}, {Michelson}, {Minuti},
  {Mitthumsiri}, {Mizuno}, {Moiseev}, {Mongelli}, {Monte}, {Monzani},
  {Moretti}, {Morselli}, {Moskalenko}, {Murgia}, {Nakajima}, {Nakamori},
  {Naumann-Godo}, {Nolan}, {Norris}, {Nuss}, {Ohno}, {Ohsugi}, {Omodei},
  {Orlando}, {Ormes}, {Ozaki}, {Paccagnella}, {Paneque}, {Panetta}, {Parent},
  {Pelassa}, {Pepe}, {Pesce-Rollins}, {Pinchera}, {Piron}, {Porter}, {Poupard},
  {Rain{\`o}}, {Rando}, {Ray}, {Razzano}, {Razzaque}, {Rea}, {Reimer},
  {Reimer}, {Reposeur}, {Ripken}, {Ritz}, {Rochester}, {Rodriguez}, {Romani},
  {Roth}, {Sadrozinski}, {Salvetti}, {Sanchez}, {Sander}, {Saz Parkinson},
  {Scargle}, {Schalk}, {Scolieri}, {Sgr{\`o}}, {Shaw}, {Siskind}, {Smith},
  {Smith}, {Spandre}, {Spinelli}, {Starck}, {Stephens}, {Striani}, {Strickman},
  {Strong}, {Suson}, {Tajima}, {Takahashi}, {Takahashi}, {Tanaka}, {Thayer},
  {Thayer}, {Thompson}, {Tibaldo}, {Tibolla}, {Tinebra}, {Torres}, {Tosti},
  {Tramacere}, {Uchiyama}, {Usher}, {Van Etten}, {Vasileiou}, {Vilchez},
  {Vitale}, {Waite}, {Wallace}, {Wang}, {Watters}, {Winer}, {Wood}, {Yang},
  {Ylinen}, {Ziegler}, \& {Fermi LAT Collaboration}}]{2010ApJS..188..405A}
{Abdo}, A.~A., {Ackermann}, M., {Ajello}, M., {et~al.} 2010{\natexlab{a}},
  \apjs, 188, 405, \dodoi{10.1088/0067-0049/188/2/405}

\bibitem[{{Abdo} {et~al.}(2010{\natexlab{b}}){Abdo}, {Ackermann}, {Ajello},
  {Antolini}, {Baldini}, {Ballet}, {Barbiellini}, {Bastieri}, {Bechtol},
  {Bellazzini}, {Berenji}, {Blandford}, {Bloom}, {Bonamente}, {Borgland},
  {Bouvier}, {Bregeon}, {Brez}, {Brigida}, {Bruel}, {Buehler}, {Burnett},
  {Buson}, {Caliandro}, {Cameron}, {Caraveo}, {Carrigan}, {Casandjian},
  {Cavazzuti}, {Cecchi}, {{\c{C}}elik}, {Chekhtman}, {Cheung}, {Chiang},
  {Ciprini}, {Claus}, {Cohen-Tanugi}, {Cominsky}, {Conrad}, {Costamante},
  {Cutini}, {Dermer}, {de Angelis}, {de Palma}, {Silva}, {Drell}, {Dubois},
  {Dumora}, {Farnier}, {Favuzzi}, {Fegan}, {Focke}, {Fortin}, {Frailis},
  {Fukazawa}, {Funk}, {Fusco}, {Gargano}, {Gasparrini}, {Gehrels}, {Germani},
  {Giebels}, {Giglietto}, {Giommi}, {Giordano}, {Glanzman}, {Godfrey},
  {Grenier}, {Grondin}, {Grove}, {Guiriec}, {Hadasch}, {Hayashida}, {Hays},
  {Healey}, {Horan}, {Hughes}, {Itoh}, {J{\'o}hannesson}, {Johnson}, {Johnson},
  {Kamae}, {Katagiri}, {Kataoka}, {Kawai}, {Kn{\"o}dlseder}, {Kuss}, {Lande},
  {Larsson}, {Latronico}, {Lemoine-Goumard}, {Longo}, {Loparco}, {Lott},
  {Lovellette}, {Lubrano}, {Madejski}, {Makeev}, {Massaro}, {Mazziotta},
  {McEnery}, {Michelson}, {Mitthumsiri}, {Mizuno}, {Moiseev}, {Monte},
  {Monzani}, {Morselli}, {Moskalenko}, {Mueller}, {Murgia}, {Nolan}, {Norris},
  {Nuss}, {Ohno}, {Ohsugi}, {Omodei}, {Orlando}, {Ormes}, {Ozaki}, {Panetta},
  {Parent}, {Pelassa}, {Pepe}, {Pesce-Rollins}, {Piron}, {Porter}, {Rain{\`o}},
  {Rando}, {Razzano}, {Reimer}, {Reimer}, {Ritz}, {Rodriguez}, {Romani},
  {Roth}, {Ryde}, {Sadrozinski}, {Sander}, {Scargle}, {Sgr{\`o}}, {Shaw},
  {Smith}, {Spandre}, {Spinelli}, {Starck}, {Strickman}, {Suson}, {Takahashi},
  {Takahashi}, {Tanaka}, {Thayer}, {Thayer}, {Thompson}, {Tibaldo}, {Torres},
  {Tosti}, {Tramacere}, {Uchiyama}, {Usher}, {Vasileiou}, {Vilchez}, {Vitale},
  {Waite}, {Wallace}, {Wang}, {Winer}, {Wood}, {Yang}, {Ylinen}, \&
  {Ziegler}}]{2010ApJ...722..520A}
---. 2010{\natexlab{b}}, \apj, 722, 520, \dodoi{10.1088/0004-637X/722/1/520}

\bibitem[{{Abdollahi} {et~al.}(2020){Abdollahi}, {Acero}, {Ackermann},
  {Ajello}, {Atwood}, {Axelsson}, {Baldini}, {Ballet}, {Barbiellini},
  {Bastieri}, {Becerra Gonzalez}, {Bellazzini}, {Berretta}, {Bissaldi},
  {Blandford}, {Bloom}, {Bonino}, {Bottacini}, {Brandt}, {Bregeon}, {Bruel},
  {Buehler}, {Burnett}, {Buson}, {Cameron}, {Caputo}, {Caraveo}, {Casandjian},
  {Castro}, {Cavazzuti}, {Charles}, {Chaty}, {Chen}, {Cheung}, {Chiaro},
  {Ciprini}, {Cohen-Tanugi}, {Cominsky}, {Coronado-Bl{\'a}zquez}, {Costantin},
  {Cuoco}, {Cutini}, {D'Ammando}, {DeKlotz}, {de la Torre Luque}, {de Palma},
  {Desai}, {Digel}, {Di Lalla}, {Di Mauro}, {Di Venere}, {Dom{\'\i}nguez},
  {Dumora}, {Fana Dirirsa}, {Fegan}, {Ferrara}, {Franckowiak}, {Fukazawa},
  {Funk}, {Fusco}, {Gargano}, {Gasparrini}, {Giglietto}, {Giommi}, {Giordano},
  {Giroletti}, {Glanzman}, {Green}, {Grenier}, {Griffin}, {Grondin}, {Grove},
  {Guiriec}, {Harding}, {Hayashi}, {Hays}, {Hewitt}, {Horan},
  {J{\'o}hannesson}, {Johnson}, {Kamae}, {Kerr}, {Kocevski}, {Kovac'evic'},
  {Kuss}, {Landriu}, {Larsson}, {Latronico}, {Lemoine-Goumard}, {Li},
  {Liodakis}, {Longo}, {Loparco}, {Lott}, {Lovellette}, {Lubrano}, {Madejski},
  {Maldera}, {Malyshev}, {Manfreda}, {Marchesini}, {Marcotulli},
  {Mart{\'\i}-Devesa}, {Martin}, {Massaro}, {Mazziotta}, {McEnery}, {Mereu},
  {Meyer}, {Michelson}, {Mirabal}, {Mizuno}, {Monzani}, {Morselli},
  {Moskalenko}, {Negro}, {Nuss}, {Ojha}, {Omodei}, {Orienti}, {Orlando},
  {Ormes}, {Palatiello}, {Paliya}, {Paneque}, {Pei}, {Pe{\~n}a-Herazo},
  {Perkins}, {Persic}, {Pesce-Rollins}, {Petrosian}, {Petrov}, {Piron}, {Poon},
  {Porter}, {Principe}, {Rain{\`o}}, {Rando}, {Razzano}, {Razzaque}, {Reimer},
  {Reimer}, {Remy}, {Reposeur}, {Romani}, {Saz Parkinson}, {Schinzel},
  {Serini}, {Sgr{\`o}}, {Siskind}, {Smith}, {Spandre}, {Spinelli}, {Strong},
  {Suson}, {Tajima}, {Takahashi}, {Tak}, {Thayer}, {Thompson}, {Tibaldo},
  {Torres}, {Torresi}, {Valverde}, {Van Klaveren}, {van Zyl}, {Wood},
  {Yassine}, \& {Zaharijas}}]{2020ApJS..247...33A}
{Abdollahi}, S., {Acero}, F., {Ackermann}, M., {et~al.} 2020, \apjs, 247, 33,
  \dodoi{10.3847/1538-4365/ab6bcb}

\bibitem[{{Abdollahi} {et~al.}(2022){Abdollahi}, {Acero}, {Baldini}, {Ballet},
  {Bastieri}, {Bellazzini}, {Berenji}, {Berretta}, {Bissaldi}, {Blandford},
  {Bloom}, {Bonino}, {Brill}, {Britto}, {Bruel}, {Burnett}, {Buson}, {Cameron},
  {Caputo}, {Caraveo}, {Castro}, {Chaty}, {Cheung}, {Chiaro}, {Cibrario},
  {Ciprini}, {Coronado-Bl{\'a}zquez}, {Crnogorcevic}, {Cutini}, {D'Ammando},
  {De Gaetano}, {Digel}, {Di Lalla}, {Dirirsa}, {Di Venere}, {Dom{\'\i}nguez},
  {Fallah Ramazani}, {Fegan}, {Ferrara}, {Fiori}, {Fleischhack}, {Franckowiak},
  {Fukazawa}, {Funk}, {Fusco}, {Galanti}, {Gammaldi}, {Gargano}, {Garrappa},
  {Gasparrini}, {Giacchino}, {Giglietto}, {Giordano}, {Giroletti}, {Glanzman},
  {Green}, {Grenier}, {Grondin}, {Guillemot}, {Guiriec}, {Gustafsson},
  {Harding}, {Hays}, {Hewitt}, {Horan}, {Hou}, {J{\'o}hannesson}, {Karwin},
  {Kayanoki}, {Kerr}, {Kuss}, {Landriu}, {Larsson}, {Latronico},
  {Lemoine-Goumard}, {Li}, {Liodakis}, {Longo}, {Loparco}, {Lott}, {Lubrano},
  {Maldera}, {Malyshev}, {Manfreda}, {Mart{\'\i}-Devesa}, {Mazziotta}, {Mereu},
  {Meyer}, {Michelson}, {Mirabal}, {Mitthumsiri}, {Mizuno}, {Moiseev},
  {Monzani}, {Morselli}, {Moskalenko}, {Negro}, {Nuss}, {Omodei}, {Orienti},
  {Orlando}, {Paneque}, {Pei}, {Perkins}, {Persic}, {Pesce-Rollins},
  {Petrosian}, {Pillera}, {Poon}, {Porter}, {Principe}, {Rain{\`o}}, {Rando},
  {Rani}, {Razzano}, {Razzaque}, {Reimer}, {Reimer}, {Reposeur},
  {S{\'a}nchez-Conde}, {Saz Parkinson}, {Scotton}, {Serini}, {Sgr{\`o}},
  {Siskind}, {Smith}, {Spandre}, {Spinelli}, {Sueoka}, {Suson}, {Tajima},
  {Tak}, {Thayer}, {Thompson}, {Torres}, {Troja}, {Valverde}, {Wood}, \&
  {Zaharijas}}]{2022ApJS..260...53A}
{Abdollahi}, S., {Acero}, F., {Baldini}, L., {et~al.} 2022, \apjs, 260, 53,
  \dodoi{10.3847/1538-4365/ac6751}

\bibitem[{{Ackermann} {et~al.}(2016){Ackermann}, {Anantua}, {Asano}, {Baldini},
  {Barbiellini}, {Bastieri}, {Becerra Gonzalez}, {Bellazzini}, {Bissaldi},
  {Blandford}, {Bloom}, {Bonino}, {Bottacini}, {Bruel}, {Buehler}, {Caliandro},
  {Cameron}, {Caragiulo}, {Caraveo}, {Cavazzuti}, {Cecchi}, {Cheung}, {Chiang},
  {Chiaro}, {Ciprini}, {Cohen-Tanugi}, {Costanza}, {Cutini}, {D'Ammando}, {de
  Palma}, {Desiante}, {Digel}, {Di Lalla}, {Di Mauro}, {Di Venere}, {Drell},
  {Favuzzi}, {Fegan}, {Ferrara}, {Fukazawa}, {Funk}, {Fusco}, {Gargano},
  {Gasparrini}, {Giglietto}, {Giordano}, {Giroletti}, {Grenier}, {Guillemot},
  {Guiriec}, {Hayashida}, {Hays}, {Horan}, {J{\'o}hannesson}, {Kensei},
  {Kocevski}, {Kuss}, {La Mura}, {Larsson}, {Latronico}, {Li}, {Longo},
  {Loparco}, {Lott}, {Lovellette}, {Lubrano}, {Madejski}, {Magill}, {Maldera},
  {Manfreda}, {Mayer}, {Mazziotta}, {Michelson}, {Mirabal}, {Mizuno},
  {Monzani}, {Morselli}, {Moskalenko}, {Nalewajko}, {Negro}, {Nuss}, {Ohsugi},
  {Orlando}, {Paneque}, {Perkins}, {Pesce-Rollins}, {Piron}, {Pivato},
  {Porter}, {Principe}, {Rando}, {Razzano}, {Razzaque}, {Reimer}, {Scargle},
  {Sgr{\`o}}, {Sikora}, {Simone}, {Siskind}, {Spada}, {Spinelli}, {Stawarz},
  {Thayer}, {Thompson}, {Torres}, {Troja}, {Uchiyama}, {Yuan}, \&
  {Zimmer}}]{2016ApJ...824L..20A}
{Ackermann}, M., {Anantua}, R., {Asano}, K., {et~al.} 2016, \apjl, 824, L20,
  \dodoi{10.3847/2041-8205/824/2/L20}

\bibitem[{{Aharonian} {et~al.}(2007){Aharonian}, {Akhperjanian}, {Bazer-Bachi},
  {Behera}, {Beilicke}, {Benbow}, {Berge}, {Bernl{\"o}hr}, {Boisson}, {Bolz},
  {Borrel}, {Boutelier}, {Braun}, {Brion}, {Brown}, {B{\"u}hler},
  {B{\"u}sching}, {Bulik}, {Carrigan}, {Chadwick}, {Clapson}, {Chounet},
  {Coignet}, {Cornils}, {Costamante}, {Degrange}, {Dickinson},
  {Djannati-Ata{\"\i}}, {Domainko}, {Drury}, {Dubus}, {Dyks}, {Egberts},
  {Emmanoulopoulos}, {Espigat}, {Farnier}, {Feinstein}, {Fiasson},
  {F{\"o}rster}, {Fontaine}, {Funk}, {Funk}, {F{\"u}{\ss}ling}, {Gallant},
  {Giebels}, {Glicenstein}, {Gl{\"u}ck}, {Goret}, {Hadjichristidis}, {Hauser},
  {Hauser}, {Heinzelmann}, {Henri}, {Hermann}, {Hinton}, {Hoffmann}, {Hofmann},
  {Holleran}, {Hoppe}, {Horns}, {Jacholkowska}, {de Jager}, {Kendziorra},
  {Kerschhaggl}, {Kh{\'e}lifi}, {Komin}, {Kosack}, {Lamanna}, {Latham}, {Le
  Gallou}, {Lemi{\`e}re}, {Lemoine-Goumard}, {Lenain}, {Lohse}, {Martin},
  {Martineau-Huynh}, {Marcowith}, {Masterson}, {Maurin}, {McComb}, {Moderski},
  {Moulin}, {de Naurois}, {Nedbal}, {Nolan}, {Olive}, {Orford}, {Osborne},
  {Ostrowski}, {Panter}, {Pedaletti}, {Pelletier}, {Petrucci}, {Pita},
  {P{\"u}hlhofer}, {Punch}, {Ranchon}, {Raubenheimer}, {Raue}, {Rayner},
  {Renaud}, {Ripken}, {Rob}, {Rolland}, {Rosier-Lees}, {Rowell}, {Rudak},
  {Ruppel}, {Sahakian}, {Santangelo}, {Saug{\'e}}, {Schlenker}, {Schlickeiser},
  {Schr{\"o}der}, {Schwanke}, {Schwarzburg}, {Schwemmer}, {Shalchi}, {Sol},
  {Spangler}, {Stawarz}, {Steenkamp}, {Stegmann}, {Superina}, {Tam},
  {Tavernet}, {Terrier}, {van Eldik}, {Vasileiadis}, {Venter}, {Vialle},
  {Vincent}, {Vivier}, {V{\"o}lk}, {Volpe}, {Wagner}, {Ward}, \&
  {Zdziarski}}]{2007ApJ...664L..71A}
{Aharonian}, F., {Akhperjanian}, A.~G., {Bazer-Bachi}, A.~R., {et~al.} 2007,
  \apjl, 664, L71, \dodoi{10.1086/520635}

\bibitem[{{Alexander}(1997)}]{1997ASSL..218..163A}
{Alexander}, T. 1997, in Astrophysics and Space Science Library, Vol. 218,
  Astronomical Time Series, ed. D.~{Maoz}, A.~{Sternberg}, \& E.~M.
  {Leibowitz}, 163, \dodoi{10.1007/978-94-015-8941-3_14}

\bibitem[{{Ansoldi} {et~al.}(2018){Ansoldi}, {Antonelli}, {Arcaro}, {Baack},
  {Babi{\'c}}, {Banerjee}, {Bangale}, {Barres de Almeida}, {Barrio}, {Becerra
  Gonz{\'a}lez}, {Bednarek}, {Bernardini}, {Berse}, {Berti}, {Besenrieder},
  {Bhattacharyya}, {Bigongiari}, {Biland}, {Blanch}, {Bonnoli}, {Carosi},
  {Ceribella}, {Chatterjee}, {Colak}, {Colin}, {Colombo}, {Contreras},
  {Cortina}, {Covino}, {Cumani}, {D'Elia}, {Da Vela}, {Dazzi}, {De Angelis},
  {De Lotto}, {Delfino}, {Delgado}, {Di Pierro}, {Dom{\'\i}nguez}, {Dominis
  Prester}, {Dorner}, {Doro}, {Einecke}, {Elsaesser}, {Fallah Ramazani},
  {Fattorini}, {Fern{\'a}ndez-Barral}, {Ferrara}, {Fidalgo}, {Foffano},
  {Fonseca}, {Font}, {Fruck}, {Gallozzi}, {Garc{\'\i}a L{\'o}pez},
  {Garczarczyk}, {Gaug}, {Giammaria}, {Godinovi{\'c}}, {Guberman}, {Hadasch},
  {Hahn}, {Hassan}, {Hayashida}, {Herrera}, {Hoang}, {Hrupec}, {Inoue},
  {Ishio}, {Iwamura}, {Konno}, {Kubo}, {Kushida}, {Lamastra}, {Lelas}, {Leone},
  {Lindfors}, {Lombardi}, {Longo}, {L{\'o}pez}, {Maggio}, {Majumdar},
  {Makariev}, {Maneva}, {Manganaro}, {Mannheim}, {Maraschi}, {Mariotti},
  {Mart{\'\i}nez}, {Masuda}, {Mazin}, {Mielke}, {Minev}, {Miranda}, {Mirzoyan},
  {Moralejo}, {Moreno}, {Moretti}, {Neustroev}, {Niedzwiecki}, {Nievas
  Rosillo}, {Nigro}, {Nilsson}, {Ninci}, {Nishijima}, {Noda}, {Nogu{\'e}s},
  {Paiano}, {Palacio}, {Paneque}, {Paoletti}, {Paredes}, {Pedaletti},
  {Pe{\~n}il}, {Peresano}, {Persic}, {Pfrang}, {Prada Moroni}, {Prandini},
  {Puljak}, {Garcia}, {Rhode}, {Rib{\'o}}, {Rico}, {Righi}, {Rugliancich},
  {Saha}, {Saito}, {Satalecka}, {Schweizer}, {Sitarek}, {{\v{S}}nidari{\'c}},
  {Sobczynska}, {Stamerra}, {Strzys}, {Suri{\'c}}, {Tavecchio}, {Temnikov},
  {Terzi{\'c}}, {Teshima}, {Torres-Alb{\'a}}, {Tsujimoto}, {Vanzo}, {Vazquez
  Acosta}, {Vovk}, {Ward}, {Will}, {Zari{\'c}}, \&
  {Cerruti}}]{2018ApJ...863L..10A}
{Ansoldi}, S., {Antonelli}, L.~A., {Arcaro}, C., {et~al.} 2018, \apjl, 863,
  L10, \dodoi{10.3847/2041-8213/aad083}

\bibitem[{{Arnaud}(1996)}]{1996ASPC..101...17A}
{Arnaud}, K.~A. 1996, in Astronomical Society of the Pacific Conference Series,
  Vol. 101, Astronomical Data Analysis Software and Systems V, ed. G.~H.
  {Jacoby} \& J.~{Barnes}, 17

\bibitem[{{Atwood} {et~al.}(2009){Atwood}, {Abdo}, {Ackermann}, {Althouse},
  {Anderson}, {Axelsson}, {Baldini}, {Ballet}, {Band}, {Barbiellini},
  {Bartelt}, {Bastieri}, {Baughman}, {Bechtol}, {B{\'e}d{\'e}r{\`e}de},
  {Bellardi}, {Bellazzini}, {Berenji}, {Bignami}, {Bisello}, {Bissaldi},
  {Blandford}, {Bloom}, {Bogart}, {Bonamente}, {Bonnell}, {Borgland},
  {Bouvier}, {Bregeon}, {Brez}, {Brigida}, {Bruel}, {Burnett}, {Busetto},
  {Caliandro}, {Cameron}, {Caraveo}, {Carius}, {Carlson}, {Casandjian},
  {Cavazzuti}, {Ceccanti}, {Cecchi}, {Charles}, {Chekhtman}, {Cheung},
  {Chiang}, {Chipaux}, {Cillis}, {Ciprini}, {Claus}, {Cohen-Tanugi},
  {Condamoor}, {Conrad}, {Corbet}, {Corucci}, {Costamante}, {Cutini}, {Davis},
  {Decotigny}, {DeKlotz}, {Dermer}, {de Angelis}, {Digel}, {do Couto e Silva},
  {Drell}, {Dubois}, {Dumora}, {Edmonds}, {Fabiani}, {Farnier}, {Favuzzi},
  {Flath}, {Fleury}, {Focke}, {Funk}, {Fusco}, {Gargano}, {Gasparrini},
  {Gehrels}, {Gentit}, {Germani}, {Giebels}, {Giglietto}, {Giommi}, {Giordano},
  {Glanzman}, {Godfrey}, {Grenier}, {Grondin}, {Grove}, {Guillemot}, {Guiriec},
  {Haller}, {Harding}, {Hart}, {Hays}, {Healey}, {Hirayama}, {Hjalmarsdotter},
  {Horn}, {Hughes}, {J{\'o}hannesson}, {Johansson}, {Johnson}, {Johnson},
  {Johnson}, {Johnson}, {Kamae}, {Katagiri}, {Kataoka}, {Kavelaars}, {Kawai},
  {Kelly}, {Kerr}, {Klamra}, {Kn{\"o}dlseder}, {Kocian}, {Komin}, {Kuehn},
  {Kuss}, {Landriu}, {Latronico}, {Lee}, {Lee}, {Lemoine-Goumard}, {Lionetto},
  {Longo}, {Loparco}, {Lott}, {Lovellette}, {Lubrano}, {Madejski}, {Makeev},
  {Marangelli}, {Massai}, {Mazziotta}, {McEnery}, {Menon}, {Meurer},
  {Michelson}, {Minuti}, {Mirizzi}, {Mitthumsiri}, {Mizuno}, {Moiseev},
  {Monte}, {Monzani}, {Moretti}, {Morselli}, {Moskalenko}, {Murgia},
  {Nakamori}, {Nishino}, {Nolan}, {Norris}, {Nuss}, {Ohno}, {Ohsugi}, {Omodei},
  {Orlando}, {Ormes}, {Paccagnella}, {Paneque}, {Panetta}, {Parent}, {Pearce},
  {Pepe}, {Perazzo}, {Pesce-Rollins}, {Picozza}, {Pieri}, {Pinchera}, {Piron},
  {Porter}, {Poupard}, {Rain{\`o}}, {Rando}, {Rapposelli}, {Razzano}, {Reimer},
  {Reimer}, {Reposeur}, {Reyes}, {Ritz}, {Rochester}, {Rodriguez}, {Romani},
  {Roth}, {Russell}, {Ryde}, {Sabatini}, {Sadrozinski}, {Sanchez}, {Sander},
  {Sapozhnikov}, {Parkinson}, {Scargle}, {Schalk}, {Scolieri}, {Sgr{\`o}},
  {Share}, {Shaw}, {Shimokawabe}, {Shrader}, {Sierpowska-Bartosik}, {Siskind},
  {Smith}, {Smith}, {Spandre}, {Spinelli}, {Starck}, {Stephens}, {Strickman},
  {Strong}, {Suson}, {Tajima}, {Takahashi}, {Takahashi}, {Tanaka}, {Tenze},
  {Tether}, {Thayer}, {Thayer}, {Thompson}, {Tibaldo}, {Tibolla}, {Torres},
  {Tosti}, {Tramacere}, {Turri}, {Usher}, {Vilchez}, {Vitale}, {Wang},
  {Watters}, {Winer}, {Wood}, {Ylinen}, \& {Ziegler}}]{2009ApJ...697.1071A}
{Atwood}, W.~B., {Abdo}, A.~A., {Ackermann}, M., {et~al.} 2009, \apj, 697,
  1071, \dodoi{10.1088/0004-637X/697/2/1071}

\bibitem[{{Begelman} {et~al.}(1984){Begelman}, {Blandford}, \&
  {Rees}}]{1984RvMP...56..255B}
{Begelman}, M.~C., {Blandford}, R.~D., \& {Rees}, M.~J. 1984, Reviews of Modern
  Physics, 56, 255, \dodoi{10.1103/RevModPhys.56.255}

\bibitem[{{Biteau} \& {Giebels}(2012)}]{2012A&A...548A.123B}
{Biteau}, J., \& {Giebels}, B. 2012, \aap, 548, A123,
  \dodoi{10.1051/0004-6361/201220056}

\bibitem[{{B{\l}a{\.z}ejowski} {et~al.}(2000){B{\l}a{\.z}ejowski}, {Sikora},
  {Moderski}, \& {Madejski}}]{2000ApJ...545..107B}
{B{\l}a{\.z}ejowski}, M., {Sikora}, M., {Moderski}, R., \& {Madejski}, G.~M.
  2000, \apj, 545, 107, \dodoi{10.1086/317791}

\bibitem[{{B{\"o}ttcher} {et~al.}(2013){B{\"o}ttcher}, {Reimer}, {Sweeney}, \&
  {Prakash}}]{2013ApJ...768...54B}
{B{\"o}ttcher}, M., {Reimer}, A., {Sweeney}, K., \& {Prakash}, A. 2013, \apj,
  768, 54, \dodoi{10.1088/0004-637X/768/1/54}

\bibitem[{{Breeveld} {et~al.}(2010){Breeveld}, {Curran}, {Hoversten}, {Koch},
  {Landsman}, {Marshall}, {Page}, {Poole}, {Roming}, {Smith}, {Still},
  {Yershov}, {Blustin}, {Brown}, {Gronwall}, {Holland}, {Kuin}, {McGowan},
  {Rosen}, {Boyd}, {Broos}, {Carter}, {Chester}, {Hancock}, {Huckle}, {Immler},
  {Ivanushkina}, {Kennedy}, {Mason}, {Morgan}, {Oates}, {de Pasquale},
  {Schady}, {Siegel}, \& {vanden Berk}}]{2010MNRAS.406.1687B}
{Breeveld}, A.~A., {Curran}, P.~A., {Hoversten}, E.~A., {et~al.} 2010, \mnras,
  406, 1687, \dodoi{10.1111/j.1365-2966.2010.16832.x}

\bibitem[{{Bulgarelli} {et~al.}(2022){Bulgarelli}, {Di Piano}, {Pittori},
  {Piano}, {Panebianco}, {Verrecchia}, {Tavani}, {Fioretti}, {Parmiggiani},
  {Addis}, {Baroncelli}, {Di Piano}, {Lucarelli}, {Vercellone}, {Cardillo},
  {Ursi}, {Casentini}, {Donnarumma}, {Gianotti}, {Trifoglio}, {Giuliani},
  {Mereghetti}, {Caraveo}, {Perotti}, {Chen}, {Argan}, {Costa}, {Del Monte},
  {Evangelista}, {Feroci}, {Foffano}, {Lapshov}, {Menegoni}, {Pacciani},
  {Soffitta}, {Vittorini}, {Lazzarotto}, {Di Cocco}, {Fuschino}, {Galli},
  {Labanti}, {Marisaldi}, {Pellizzoni}, {Pilia}, {Trois}, {Barbiellini},
  {Longo}, {Vallazza}, {Morselli}, {Picozza}, {Prest}, {Lipari}, {Zanello},
  {Cattaneo}, {Rappoldi}, {Ferrari}, {Paoletti}, {Antonelli}, {Giommi},
  {Salotti}, {Valentini}, \& {D'Amico}}]{2022ATel15768....1B}
{Bulgarelli}, A., {Di Piano}, A., {Pittori}, C., {et~al.} 2022, The
  Astronomer's Telegram, 15768, 1

\bibitem[{{Ciprini}(2009{\natexlab{a}})}]{2009ATel.2136....1C}
{Ciprini}, S. 2009{\natexlab{a}}, The Astronomer's Telegram, 2136, 1

\bibitem[{{Ciprini}(2009{\natexlab{b}})}]{2009ATel.2048....1C}
---. 2009{\natexlab{b}}, The Astronomer's Telegram, 2048, 1

\bibitem[{{Dermer}(1995)}]{1995ApJ...446L..63D}
{Dermer}, C.~D. 1995, \apjl, 446, L63, \dodoi{10.1086/187931}

\bibitem[{{Dermer} {et~al.}(1992){Dermer}, {Schlickeiser}, \&
  {Mastichiadis}}]{1992A&A...256L..27D}
{Dermer}, C.~D., {Schlickeiser}, R., \& {Mastichiadis}, A. 1992, \aap, 256, L27

\bibitem[{{Di Gesu} {et~al.}(2022){Di Gesu}, {Donnarumma}, {Tavecchio},
  {Agudo}, {Barnounin}, {Cibrario}, {Di Lalla}, {Di Marco}, {Escudero},
  {Errando}, {Jorstad}, {Kim}, {Kouch}, {Liodakis}, {Lindfors}, {Madejski},
  {Marshall}, {Marscher}, {Middei}, {Muleri}, {Myserlis}, {Negro}, {Omodei},
  {Pacciani}, {Paggi}, {Perri}, {Puccetti}, {Antonelli}, {Bachetti}, {Baldini},
  {Baumgartner}, {Bellazzini}, {Bianchi}, {Bongiorno}, {Bonino}, {Brez},
  {Bucciantini}, {Capitanio}, {Castellano}, {Cavazzuti}, {Ciprini}, {Costa},
  {De Rosa}, {Del Monte}, {Doroshenko}, {Dov{\v{c}}iak}, {Ehlert}, {Enoto},
  {Evangelista}, {Fabiani}, {Ferrazzoli}, {Garcia}, {Gunji}, {Hayashida},
  {Heyl}, {Iwakiri}, {Karas}, {Kitaguchi}, {Kolodziejczak}, {Krawczynski}, {La
  Monaca}, {Latronico}, {Maldera}, {Manfreda}, {Marin}, {Marinucci}, {Massaro},
  {Matt}, {Mitsuishi}, {Mizuno}, {Ng}, {O'Dell}, {Oppedisano}, {Papitto},
  {Pavlov}, {Peirson}, {Pesce-Rollins}, {Petrucci}, {Pilia}, {Possenti},
  {Poutanen}, {Ramsey}, {Rankin}, {Ratheesh}, {Romani}, {Sgr{\`o}}, {Slane},
  {Soffitta}, {Spandre}, {Tamagawa}, {Taverna}, {Tawara}, {Tennant}, {Thomas},
  {Tombesi}, {Trois}, {Tsygankov}, {Turolla}, {Vink}, {Weisskopf}, {Wu}, {Xie},
  \& {Zane}}]{2022ApJ...938L...7D}
{Di Gesu}, L., {Donnarumma}, I., {Tavecchio}, F., {et~al.} 2022, \apjl, 938,
  L7, \dodoi{10.3847/2041-8213/ac913a}

\bibitem[{{Edwards} {et~al.}(2022){Edwards}, {Stevens}, {Kadler}, {Ojha},
  {Tornikoski}, \& {Lahteenmaki}}]{2022ATel15692....1E}
{Edwards}, P.~G., {Stevens}, J., {Kadler}, M., {et~al.} 2022, The Astronomer's
  Telegram, 15692, 1

\bibitem[{{Eisenstein} \& {Hut}(1998)}]{1998ApJ...498..137E}
{Eisenstein}, D.~J., \& {Hut}, P. 1998, \apj, 498, 137, \dodoi{10.1086/305535}

\bibitem[{{Fan} {et~al.}(2011){Fan}, {Xu}, {Pan}, \&
  {Yuan}}]{2011IAUS..275..164F}
{Fan}, J.~H., {Xu}, W., {Pan}, J., \& {Yuan}, Y.~H. 2011, in Jets at All
  Scales, ed. G.~E. {Romero}, R.~A. {Sunyaev}, \& T.~{Belloni}, Vol. 275,
  164--167, \dodoi{10.1017/S1743921310015875}

\bibitem[{Fan {et~al.}(2021)Fan, Kurtanidze, Liu, Kurtanidze, Nikolashvili,
  Liu, Zhang, Cai, Zhu, He, Yang, Yang, Gu, Luo, \& Yuan}]{Fan_2021}
Fan, J.~H., Kurtanidze, S.~O., Liu, Y., {et~al.} 2021, The Astrophysical
  Journal Supplement Series, 253, 10, \dodoi{10.3847/1538-4365/abd32d}

\bibitem[{{Finke} {et~al.}(2008){Finke}, {Dermer}, \&
  {B{\"o}ttcher}}]{2008ApJ...686..181F}
{Finke}, J.~D., {Dermer}, C.~D., \& {B{\"o}ttcher}, M. 2008, \apj, 686, 181,
  \dodoi{10.1086/590900}

\bibitem[{{Fitzpatrick} \& {Massa}(2007)}]{2007ApJ...663..320F}
{Fitzpatrick}, E.~L., \& {Massa}, D. 2007, \apj, 663, 320,
  \dodoi{10.1086/518158}

\bibitem[{{Foschini} {et~al.}(2011){Foschini}, {Ghisellini}, {Tavecchio},
  {Bonnoli}, \& {Stamerra}}]{2011A&A...530A..77F}
{Foschini}, L., {Ghisellini}, G., {Tavecchio}, F., {Bonnoli}, G., \&
  {Stamerra}, A. 2011, \aap, 530, A77, \dodoi{10.1051/0004-6361/201117064}

\bibitem[{{Gasparyan} {et~al.}(2022){Gasparyan}, {B{\'e}gu{\'e}}, \&
  {Sahakyan}}]{2022MNRAS.509.2102G}
{Gasparyan}, S., {B{\'e}gu{\'e}}, D., \& {Sahakyan}, N. 2022, \mnras, 509,
  2102, \dodoi{10.1093/mnras/stab2688}

\bibitem[{{Gehrels} {et~al.}(2004){Gehrels}, {Chincarini}, {Giommi}, {Mason},
  {Nousek}, {Wells}, {White}, {Barthelmy}, {Burrows}, {Cominsky}, {Hurley},
  {Marshall}, {M{\'e}sz{\'a}ros}, {Roming}, {Angelini}, {Barbier}, {Belloni},
  {Campana}, {Caraveo}, {Chester}, {Citterio}, {Cline}, {Cropper}, {Cummings},
  {Dean}, {Feigelson}, {Fenimore}, {Frail}, {Fruchter}, {Garmire}, {Gendreau},
  {Ghisellini}, {Greiner}, {Hill}, {Hunsberger}, {Krimm}, {Kulkarni}, {Kumar},
  {Lebrun}, {Lloyd-Ronning}, {Markwardt}, {Mattson}, {Mushotzky}, {Norris},
  {Osborne}, {Paczynski}, {Palmer}, {Park}, {Parsons}, {Paul}, {Rees},
  {Reynolds}, {Rhoads}, {Sasseen}, {Schaefer}, {Short}, {Smale}, {Smith},
  {Stella}, {Tagliaferri}, {Takahashi}, {Tashiro}, {Townsley}, {Tueller},
  {Turner}, {Vietri}, {Voges}, {Ward}, {Willingale}, {Zerbi}, \&
  {Zhang}}]{2004ApJ...611.1005G}
{Gehrels}, N., {Chincarini}, G., {Giommi}, P., {et~al.} 2004, \apj, 611, 1005,
  \dodoi{10.1086/422091}

\bibitem[{{Ghisellini} {et~al.}(1993){Ghisellini}, {Padovani}, {Celotti}, \&
  {Maraschi}}]{1993ApJ...407...65G}
{Ghisellini}, G., {Padovani}, P., {Celotti}, A., \& {Maraschi}, L. 1993, \apj,
  407, 65, \dodoi{10.1086/172493}

\bibitem[{{Giebels} \& {Degrange}(2009)}]{2009A&A...503..797G}
{Giebels}, B., \& {Degrange}, B. 2009, \aap, 503, 797,
  \dodoi{10.1051/0004-6361/200912303}

\bibitem[{{Goswami} {et~al.}(2018){Goswami}, {Sahayanathan}, {Sinha}, {Misra},
  \& {Gogoi}}]{2018MNRAS.480.2046G}
{Goswami}, P., {Sahayanathan}, S., {Sinha}, A., {Misra}, R., \& {Gogoi}, R.
  2018, \mnras, 480, 2046, \dodoi{10.1093/mnras/sty2003}

\bibitem[{{Hota} {et~al.}(2021){Hota}, {Shah}, {Khatoon}, {Misra}, {Pradhan},
  \& {Gogoi}}]{2021MNRAS.508.5921H}
{Hota}, J., {Shah}, Z., {Khatoon}, R., {et~al.} 2021, \mnras, 508, 5921,
  \dodoi{10.1093/mnras/stab2903}

\bibitem[{{IceCube Collaboration} {et~al.}(2018){IceCube Collaboration},
  {Aartsen}, {Ackermann}, {Adams}, {Aguilar}, {Ahlers}, {Ahrens}, {Al Samarai},
  {Altmann}, {Andeen}, {Anderson}, {Ansseau}, {Anton}, {Arg{\"u}elles},
  {Auffenberg}, {Axani}, {Bagherpour}, {Bai}, {Barron}, {Barwick}, {Baum},
  {Bay}, {Beatty}, {Becker Tjus}, {Becker}, {BenZvi}, {Berley}, {Bernardini},
  {Besson}, {Binder}, {Bindig}, {Blaufuss}, {Blot}, {Bohm}, {B{\"o}rner},
  {Bos}, {B{\"o}ser}, {Botner}, {Bourbeau}, {Bourbeau}, {Bradascio}, {Braun},
  {Brenzke}, {Bretz}, {Bron}, {Brostean-Kaiser}, {Burgman}, {Busse}, {Carver},
  {Cheung}, {Chirkin}, {Christov}, {Clark}, {Classen}, {Coenders}, {Collin},
  {Conrad}, {Coppin}, {Correa}, {Cowen}, {Cross}, {Dave}, {Day}, {de
  Andr{\'e}}, {De Clercq}, {DeLaunay}, {Dembinski}, {De Ridder}, {Desiati}, {de
  Vries}, {de Wasseige}, {de With}, {DeYoung}, {D{\'\i}az-V{\'e}lez}, {di
  Lorenzo}, {Dujmovic}, {Dumm}, {Dunkman}, {Dvorak}, {Eberhardt}, {Ehrhardt},
  {Eichmann}, {Eller}, {Evenson}, {Fahey}, {Fazely}, {Felde}, {Filimonov},
  {Finley}, {Flis}, {Franckowiak}, {Friedman}, {Fritz}, {Gaisser}, {Gallagher},
  {Gerhardt}, {Ghorbani}, {Glauch}, {Gl{\"u}senkamp}, {Goldschmidt},
  {Gonzalez}, {Grant}, {Griffith}, {Haack}, {Hallgren}, {Halzen}, {Hanson},
  {Hebecker}, {Heereman}, {Helbing}, {Hellauer}, {Hickford}, {Hignight},
  {Hill}, {Hoffman}, {Hoffmann}, {Hoinka}, {Hokanson-Fasig}, {Hoshina},
  {Huang}, {Huber}, {Hultqvist}, {H{\"u}nnefeld}, {Hussain}, {In}, {Iovine},
  {Ishihara}, {Jacobi}, {Japaridze}, {Jeong}, {Jero}, {Jones}, {Kalaczynski},
  {Kang}, {Kappes}, {Kappesser}, {Karg}, {Karle}, {Katz}, {Kauer}, {Keivani},
  {Kelley}, {Kheirandish}, {Kim}, {Kim}, {Kintscher}, {Kiryluk}, {Kittler},
  {Klein}, {Koirala}, {Kolanoski}, {K{\"o}pke}, {Kopper}, {Kopper},
  {Koschinsky}, {Koskinen}, {Kowalski}, {Krings}, {Kroll}, {Kr{\"u}ckl},
  {Kunwar}, {Kurahashi}, {Kuwabara}, {Kyriacou}, {Labare}, {Lanfranchi},
  {Larson}, {Lauber}, {Leonard}, {Lesiak-Bzdak}, {Leuermann}, {Liu}, {Lozano
  Mariscal}, {Lu}, {L{\"u}nemann}, {Luszczak}, {Madsen}, {Maggi}, {Mahn},
  {Mancina}, {Maruyama}, {Mase}, {Maunu}, {Meagher}, {Medici}, {Meier},
  {Menne}, {Merino}, {Meures}, {Miarecki}, {Micallef}, {Moment{\'e}},
  {Montaruli}, {Moore}, {Morse}, {Moulai}, {Nahnhauer}, {Nakarmi}, {Naumann},
  {Neer}, {Niederhausen}, {Nowicki}, {Nygren}, {Obertacke Pollmann}, {Olivas},
  {O'Murchadha}, {O'Sullivan}, {Palczewski}, {Pandya}, {Pankova}, {Peiffer},
  {Pepper}, {P{\'e}rez de los Heros}, {Pieloth}, {Pinat}, {Plum}, {Price},
  {Przybylski}, {Raab}, {R{\"a}del}, {Rameez}, {Rauch}, {Rawlins}, {Rea},
  {Reimann}, {Relethford}, {Relich}, {Resconi}, {Rhode}, {Richman},
  {Robertson}, {Rongen}, {Rott}, {Ruhe}, {Ryckbosch}, {Rysewyk}, {Safa},
  {S{\"a}lzer}, {Sanchez Herrera}, {Sandrock}, {Sandroos}, {Santander},
  {Sarkar}, {Sarkar}, {Satalecka}, {Schlunder}, {Schmidt}, {Schneider},
  {Schoenen}, {Sch{\"o}neberg}, {Schumacher}, {Sclafani}, {Seckel},
  {Seunarine}, {Soedingrekso}, {Soldin}, {Song}, {Spiczak}, {Spiering},
  {Stachurska}, {Stamatikos}, {Stanev}, {Stasik}, {Stein}, {Stettner},
  {Steuer}, {Stezelberger}, {Stokstad}, {St{\"o}{\ss}l}, {Strotjohann},
  {Stuttard}, {Sullivan}, {Sutherland}, {Taboada}, {Tatar}, {Tenholt},
  {Ter-Antonyan}, {Terliuk}, {Tilav}, {Toale}, {Tobin}, {Toennis}, {Toscano},
  {Tosi}, {Tselengidou}, {Tung}, {Turcati}, {Turley}, {Ty}, {Unger}, {Usner},
  {Vandenbroucke}, {Van Driessche}, {van Eijk}, {van Eijndhoven}, {Vanheule},
  {van Santen}, {Vogel}, {Vraeghe}, {Walck}, {Wallace}, {Wallraff}, {Wandler},
  {Wandkowsky}, {Waza}, {Weaver}, {Weiss}, {Wendt}, {Werthebach}, {Westerhoff},
  {Whelan}, {Whitehorn}, {Wiebe}, {Wiebusch}, {Wille}, {Williams}, {Wills},
  {Wolf}, {Wood}, {Wood}, {Woschnagg}, {Xu}, {Xu}, {Xu}, {Yanez}, {Yodh},
  {Yoshida}, {Yuan}, {Fermi-LAT Collaboration}, {Abdollahi}, {Ajello},
  {Angioni}, {Baldini}, {Ballet}, {Barbiellini}, {Bastieri}, {Bechtol},
  {Bellazzini}, {Berenji}, {Bissaldi}, {Blandford}, {Bonino}, {Bottacini},
  {Bregeon}, {Bruel}, {Buehler}, {Burnett}, {Burns}, {Buson}, {Cameron},
  {Caputo}, {Caraveo}, {Cavazzuti}, {Charles}, {Chen}, {Cheung}, {Chiang},
  {Chiaro}, {Ciprini}, {Cohen-Tanugi}, {Conrad}, {Costantin}, {Cutini},
  {D'Ammando}, {de Palma}, {Digel}, {Di Lalla}, {Di Mauro}, {Di Venere},
  {Dom{\'\i}nguez}, {Favuzzi}, {Franckowiak}, {Fukazawa}, {Funk}, {Fusco},
  {Gargano}, {Gasparrini}, {Giglietto}, {Giomi}, {Giommi}, {Giordano},
  {Giroletti}, {Glanzman}, {Green}, {Grenier}, {Grondin}, {Guiriec}, {Harding},
  {Hayashida}, {Hays}, {Hewitt}, {Horan}, {J{\'o}hannesson}, {Kadler},
  {Kensei}, {Kocevski}, {Krauss}, {Kreter}, {Kuss}, {La Mura}, {Larsson},
  {Latronico}, {Lemoine-Goumard}, {Li}, {Longo}, {Loparco}, {Lovellette},
  {Lubrano}, {Magill}, {Maldera}, {Malyshev}, {Manfreda}, {Mazziotta},
  {McEnery}, {Meyer}, {Michelson}, {Mizuno}, {Monzani}, {Morselli},
  {Moskalenko}, {Negro}, {Nuss}, {Ojha}, {Omodei}, {Orienti}, {Orlando},
  {Palatiello}, {Paliya}, {Perkins}, {Persic}, {Pesce-Rollins}, {Piron},
  {Porter}, {Principe}, {Rain{\`o}}, {Rando}, {Rani}, {Razzano}, {Razzaque},
  {Reimer}, {Reimer}, {Renault-Tinacci}, {Ritz}, {Rochester}, {Saz Parkinson},
  {Sgr{\`o}}, {Siskind}, {Spandre}, {Spinelli}, {Suson}, {Tajima}, {Takahashi},
  {Tanaka}, {Thayer}, {Thompson}, {Tibaldo}, {Torres}, {Torresi}, {Tosti},
  {Troja}, {Valverde}, {Vianello}, {Vogel}, {Wood}, {Wood}, {Zaharijas}, {MAGIC
  Collaboration}, {Ahnen}, {Ansoldi}, {Antonelli}, {Arcaro}, {Baack},
  {Babi{\'c}}, {Banerjee}, {Bangale}, {Barres de Almeida}, {Barrio}, {Becerra
  Gonz{\'a}lez}, {Bednarek}, {Bernardini}, {Berti}, {Bhattacharyya}, {Biland},
  {Blanch}, {Bonnoli}, {Carosi}, {Carosi}, {Ceribella}, {Chatterjee}, {Colak},
  {Colin}, {Colombo}, {Contreras}, {Cortina}, {Covino}, {Cumani}, {Da Vela},
  {Dazzi}, {De Angelis}, {De Lotto}, {Delfino}, {Delgado}, {Di Pierro},
  {Dom{\'\i}nguez}, {Dominis Prester}, {Dorner}, {Doro}, {Einecke},
  {Elsaesser}, {Fallah Ramazani}, {Fern{\'a}ndez-Barral}, {Fidalgo}, {Foffano},
  {Pfrang}, {Fonseca}, {Font}, {Franceschini}, {Fruck}, {Galindo}, {Gallozzi},
  {Garc{\'\i}a L{\'o}pez}, {Garczarczyk}, {Gaug}, {Giammaria}, {Godinovi{\'c}},
  {Gora}, {Guberman}, {Hadasch}, {Hahn}, {Hassan}, {Hayashida}, {Herrera},
  {Hose}, {Hrupec}, {Inoue}, {Ishio}, {Konno}, {Kubo}, {Kushida}, {Lelas},
  {Lindfors}, {Lombardi}, {Longo}, {L{\'o}pez}, {Maggio}, {Majumdar},
  {Makariev}, {Maneva}, {Manganaro}, {Mannheim}, {Maraschi}, {Mariotti},
  {Mart{\'\i}nez}, {Masuda}, {Mazin}, {Minev}, {M}, {Mirzoyan}, {Moralejo},
  {Moreno}, {Moretti}, {Nagayoshi}, {Neustroev}, {Niedzwiecki}, {Nievas
  Rosillo}, {Nigro}, {Nilsson}, {Ninci}, {Nishijima}, {Noda}, {Nogu{\'e}s},
  {Paiano}, {Palacio}, {Paneque}, {Paoletti}, {Paredes}, {Pedaletti},
  {Peresano}, {Persic}, {Prada Moroni}, {Prandini}, {Puljak}, {Rodriguez
  Garcia}, {Reichardt}, {Rhode}, {Rib{\'o}}, {Rico}, {Righi}, {Rugliancich},
  {Saito}, {Satalecka}, {Schweizer}, {Sitarek}, {{\v{S}}nidaric
  {\textasciiacute}}, {Sobczynska}, {Stamerra}, {Strzys}, {Suri{\'c}},
  {Takahashi}, {Tavecchio}, {Temnikov}, {Terzi{\'c}}, {Teshima},
  {Torres-Alb{\`a}}, {Treves}, {Tsujimoto}, {Vanzo}, {Vazquez Acosta}, {Vovk},
  {Ward}, {Will}, {S}, {Zaric {\textasciiacute}}, {AGILE Team}, {Lucarelli},
  {Tavani}, {Piano}, {Donnarumma}, {Pittori}, {Verrecchia}, {Barbiellini},
  {Bulgarelli}, {Caraveo}, {Cattaneo}, {Colafrancesco}, {Costa}, {Di Cocco},
  {Ferrari}, {Gianotti}, {Giuliani}, {Lipari}, {Mereghetti}, {Morselli},
  {Pacciani}, {Paoletti}, {Parmiggiani}, {Pellizzoni}, {Picozza}, {Pilia},
  {Rappoldi}, {Trois}, {Vercellone}, {Vittorini}, {ASAS-SN Team}, {Stanek},
  {Franckowiak}, {Kochanek}, {Beacom}, {Thompson}, {Holoien}, {Dong}, {Prieto},
  {Shappee}, {Holmbo}, {HAWC Collaboration}, {Abeysekara}, {Albert}, {Alfaro},
  {Alvarez}, {Arceo}, {Arteaga-Vel{\'a}zquez}, {Avila Rojas}, {Ayala Solares},
  {Becerril}, {Belmont-Moreno}, {Bernal}, {Caballero-Mora}, {Capistr{\'a}n},
  {Carrami{\~n}ana}, {Casanova}, {Castillo}, {Cotti}, {Cotzomi}, {Couti{\~n}o
  de Le{\'o}n}, {De Le{\'o}n}, {De la Fuente}, {Diaz Hernandez}, {Dichiara},
  {Dingus}, {DuVernois}, {D{\'\i}az-V{\'e}lez}, {Ellsworth}, {Engel},
  {Fiorino}, {Fleischhack}, {Fraija}, {Garc{\'\i}a-Gonz{\'a}lez}, {Garfias},
  {Gonz{\'a}lez Mu{\~n}oz}, {Gonz{\'a}lez}, {Goodman}, {Hampel-Arias},
  {Harding}, {Hernandez}, {Hona}, {Hueyotl-Zahuantitla}, {Hui},
  {H{\"u}ntemeyer}, {Iriarte}, {Jardin-Blicq}, {Joshi}, {Kaufmann}, {Kunde},
  {Lara}, {Lauer}, {Lee}, {Lennarz}, {Le{\'o}n Vargas}, {Linnemann},
  {Longinotti}, {Luis-Raya}, {Luna-Garc{\'\i}a}, {Malone}, {Marinelli},
  {Martinez}, {Martinez-Castellanos}, {Mart{\'\i}nez-Castro},
  {Mart{\'\i}nez-Huerta}, {Matthews}, {Miranda-Romagnoli}, {Moreno},
  {Mostaf{\'a}}, {Nayerhoda}, {Nellen}, {Newbold}, {Nisa}, {Noriega-Papaqui},
  {Pelayo}, {Pretz}, {P{\'e}rez-P{\'e}rez}, {Ren}, {Rho}, {Rivi{\`e}re},
  {Rosa-Gonz{\'a}lez}, {Rosenberg}, {Ruiz-Velasco}, {Ruiz-Velasco}, {Salesa
  Greus}, {Sandoval}, {Schneider}, {Schoorlemmer}, {Sinnis}, {Smith},
  {Springer}, {Surajbali}, {Tibolla}, {Tollefson}, {Torres}, {Villase{\~n}or},
  {Weisgarber}, {Werner}, {Yapici}, {Gaurang}, {Zepeda}, {Zhou}, {{\'A}lvarez},
  {H.~E.~S.~S. Collaboration}, {Abdalla}, {Ang{\"u}ner}, {Armand}, {Backes},
  {Becherini}, {Berge}, {B{\"o}ttcher}, {Boisson}, {Bolmont}, {Bonnefoy},
  {Bordas}, {Brun}, {B{\"u}chele}, {Bulik}, {Caroff}, {Carosi}, {Casanova},
  {Cerruti}, {Chakraborty}, {Chandra}, {Chen}, {Colafrancesco}, {Davids},
  {Deil}, {Devin}, {Djannati-Ata{\"\i}}, {Egberts}, {Emery}, {Eschbach},
  {Fiasson}, {Fontaine}, {Funk}, {F{\"u}{\ss}ling}, {Gallant}, {Gat{\'e}},
  {Giavitto}, {Glawion}, {Glicenstein}, {Gottschall}, {Grondin}, {Haupt},
  {Henri}, {Hinton}, {Hoischen}, {Holch}, {Huber}, {Jamrozy}, {Jankowsky},
  {Jankowsky}, {Jouvin}, {Jung-Richardt}, {Kerszberg}, {Kh{\'e}lifi}, {King},
  {Klepser}, {Kluz {\textasciiacute}niak}, {Komin}, {Kraus}, {Lefaucheur},
  {Lemi{\`e}re}, {Lemoine-Goumard}, {Lenain}, {Leser}, {Lohse},
  {L{\'o}pez-Coto}, {Lorentz}, {Lypova}, {Marandon}, {Guillem
  Mart{\'\i}-Devesa}, {Maurin}, {Mitchell}, {Moderski}, {Mohamed}, {Mohrmann},
  {Moulin}, {Murach}, {de Naurois}, {Niederwanger}, {Niemiec}, {Oakes},
  {O'Brien}, {Ohm}, {Ostrowski}, {Oya}, {Panter}, {Parsons}, {Perennes},
  {Piel}, {Pita}, {Poireau}, {Priyana Noel}, {Prokoph}, {P{\"u}hlhofer},
  {Quirrenbach}, {Raab}, {Rauth}, {Renaud}, {Rieger}, {Rinchiuso}, {Romoli},
  {Rowell}, {Rudak}, {Sasaki}, {Sanchez}, {Schlickeiser}, {Sch{\"u}ssler},
  {Schulz}, {Schwanke}, {Seglar-Arroyo}, {Shafi}, {Simoni}, {Sol}, {Stegmann},
  {Steppa}, {Tavernier}, {Taylor}, {Tiziani}, {Trichard}, {Tsirou}, {van
  Eldik}, {van Rensburg}, {van Soelen}, {Veh}, {Vincent}, {Voisin}, {Wagner},
  {Wagner}, {Wierzcholska}, {Zanin}, {Zdziarski}, {Zech}, {Ziegler}, {Zorn},
  {{\.Z}ywucka}, {INTEGRAL Team}, {Savchenko}, {Ferrigno}, {Bazzano}, {Diehl},
  {Kuulkers}, {Laurent}, {Mereghetti}, {Natalucci}, {Panessa}, {Rodi},
  {Ubertini}, {Kanata}, Teams, {Morokuma}, {Ohta}, {Tanaka}, {Mori},
  {Yamanaka}, {Kawabata}, {Utsumi}, {Nakaoka}, {Kawabata}, {Nagashima},
  {Yoshida}, {Matsuoka}, {Itoh}, {Kapteyn Team}, {Keel}, {Liverpool Telescope
  Team}, {Copperwheat}, {Steele}, {Swift/NuSTAR Team}, {Cenko}, {Cowen},
  {DeLaunay}, {Evans}, {Fox}, {Keivani}, {Kennea}, {Marshall}, {Osborne},
  {Santander}, {Tohuvavohu}, {Turley}, {VERITAS Collaboration}, {Abeysekara},
  {Archer}, {Benbow}, {Bird}, {Brill}, {Brose}, {Buchovecky}, {Buckley},
  {Bugaev}, {Christiansen}, {Connolly}, {Cui}, {Daniel}, {Errando}, {Falcone},
  {Feng}, {Finley}, {Fortson}, {Furniss}, {Gueta}, {H{\"u}tten}, {Hervet},
  {Hughes}, {Humensky}, {Johnson}, {Kaaret}, {Kar}, {Kelley-Hoskins},
  {Kertzman}, {Kieda}, {Krause}, {Krennrich}, {Kumar}, {Lang}, {Lin}, {Maier},
  {McArthur}, {Moriarty}, {Mukherjee}, {Nieto}, {O'Brien}, {Ong}, {Otte},
  {Park}, {Petrashyk}, {Pohl}, {Popkow}, {Pueschel}, {Quinn}, {Ragan},
  {Reynolds}, {Richards}, {Roache}, {Rulten}, {Sadeh}, {Santander}, {Scott},
  {Sembroski}, {Shahinyan}, {Sushch}, {Tr{\'e}panier}, {Tyler}, {Vassiliev},
  {Wakely}, {Weinstein}, {Wells}, {Wilcox}, {Wilhelm}, {Williams}, {Zitzer},
  {VLA/B Team}, {Tetarenko}, {Kimball}, {Miller-Jones}, \&
  {Sivakoff}}]{2018Sci...361.1378I}
{IceCube Collaboration}, {Aartsen}, M.~G., {Ackermann}, M., {et~al.} 2018,
  Science, 361, eaat1378, \dodoi{10.1126/science.aat1378}

\bibitem[{{Jeong} {et~al.}(2023){Jeong}, {Lee}, {Cheong}, {Kim}, {Lee}, {Kang},
  {Kim}, {Rani}, {Park}, \& {Gurwell}}]{2023MNRAS.523.5703J}
{Jeong}, H.-W., {Lee}, S.-S., {Cheong}, W.~Y., {et~al.} 2023, \mnras, 523,
  5703, \dodoi{10.1093/mnras/stad1736}

\bibitem[{{Jones} {et~al.}(1974){Jones}, {O'Dell}, \&
  {Stein}}]{1974ApJ...188..353J}
{Jones}, T.~W., {O'Dell}, S.~L., \& {Stein}, W.~A. 1974, \apj, 188, 353,
  \dodoi{10.1086/152724}

\bibitem[{{Jorstad} {et~al.}(2022){Jorstad}, {Marscher}, {Raiteri}, {Villata},
  {Weaver}, {Zhang}, {Dong}, {G{\'o}mez}, {Perel}, {Savchenko}, {Larionov},
  {Carosati}, {Chen}, {Kurtanidze}, {Marchini}, {Matsumoto}, {Mortari},
  {Aceti}, {Acosta-Pulido}, {Andreeva}, {Apolonio}, {Arena}, {Arkharov},
  {Bachev}, {Banfi}, {Bonnoli}, {Borman}, {Bozhilov}, {Carnerero},
  {Damljanovic}, {Ehgamberdiev}, {Els{\"a}sser}, {Frasca}, {Gabellini},
  {Grishina}, {Gupta}, {Hagen-Thorn}, {Hallum}, {Hart}, {Hasuda}, {Hemrich},
  {Hsiao}, {Ibryamov}, {Irsmambetova}, {Ivanov}, {Joner}, {Kimeridze},
  {Klimanov}, {Kn{\"o}tt}, {Kopatskaya}, {Kurtanidze}, {Kurtenkov}, {Kuutma},
  {Larionova}, {Leonini}, {Lin}, {Lorey}, {Mannheim}, {Marino}, {Minev},
  {Mirzaqulov}, {Morozova}, {Nikiforova}, {Nikolashvili}, {Ovcharov}, {Papini},
  {Pursimo}, {Rahimov}, {Reinhart}, {Sakamoto}, {Salvaggio}, {Semkov},
  {Shakhovskoy}, {Sigua}, {Steineke}, {Stojanovic}, {Strigachev}, {Troitskaya},
  {Troitskiy}, {Tsai}, {Valcheva}, {Vasilyev}, {Vince}, {Waller}, {Zaharieva},
  \& {Chatterjee}}]{2022Natur.609..265J}
{Jorstad}, S.~G., {Marscher}, A.~P., {Raiteri}, C.~M., {et~al.} 2022, \nat,
  609, 265, \dodoi{10.1038/s41586-022-05038-9}

\bibitem[{{Kalberla} {et~al.}(2005){Kalberla}, {Burton}, {Hartmann}, {Arnal},
  {Bajaja}, {Morras}, \& {P{\"o}ppel}}]{2005A&A...440..775K}
{Kalberla}, P.~M.~W., {Burton}, W.~B., {Hartmann}, D., {et~al.} 2005, \aap,
  440, 775, \dodoi{10.1051/0004-6361:20041864}

\bibitem[{{Keivani} {et~al.}(2018){Keivani}, {Murase}, {Petropoulou}, {Fox},
  {Cenko}, {Chaty}, {Coleiro}, {DeLaunay}, {Dimitrakoudis}, {Evans}, {Kennea},
  {Marshall}, {Mastichiadis}, {Osborne}, {Santander}, {Tohuvavohu}, \&
  {Turley}}]{2018ApJ...864...84K}
{Keivani}, A., {Murase}, K., {Petropoulou}, M., {et~al.} 2018, \apj, 864, 84,
  \dodoi{10.3847/1538-4357/aad59a}

\bibitem[{{Khatoon} {et~al.}(2022){Khatoon}, {Shah}, {Hota}, {Misra}, {Gogoi},
  \& {Pradhan}}]{2022MNRAS.515.3749K}
{Khatoon}, R., {Shah}, Z., {Hota}, J., {et~al.} 2022, \mnras, 515, 3749,
  \dodoi{10.1093/mnras/stac1964}

\bibitem[{{Kim} {et~al.}(2022){Kim}, {Lee}, {Lee}, {Hodgson}, {Kang}, {Algaba},
  {Kim}, {Hodges}, {Agudo}, {Fuentes}, {Escudero}, {Myserlis}, {Traianou},
  {L{\"a}hteenm{\"a}ki}, {Tornikoski}, {Tammi}, {Ramakrishnan}, \&
  {J{\"a}rvel{\"a}}}]{2022MNRAS.510..815K}
{Kim}, S.-H., {Lee}, S.-S., {Lee}, J.~W., {et~al.} 2022, \mnras, 510, 815,
  \dodoi{10.1093/mnras/stab3473}

\bibitem[{{Krawczynski} {et~al.}(2002){Krawczynski}, {Coppi}, \&
  {Aharonian}}]{2002MNRAS.336..721K}
{Krawczynski}, H., {Coppi}, P.~S., \& {Aharonian}, F. 2002, \mnras, 336, 721,
  \dodoi{10.1046/j.1365-8711.2002.05750.x}

\bibitem[{{La Mura}(2022)}]{2022ATel15676....1L}
{La Mura}, G. 2022, The Astronomer's Telegram, 15676, 1

\bibitem[{{Lister} {et~al.}(2019){Lister}, {Homan}, {Hovatta}, {Kellermann},
  {Kiehlmann}, {Kovalev}, {Max-Moerbeck}, {Pushkarev}, {Readhead}, {Ros}, \&
  {Savolainen}}]{2019ApJ...874...43L}
{Lister}, M.~L., {Homan}, D.~C., {Hovatta}, T., {et~al.} 2019, \apj, 874, 43,
  \dodoi{10.3847/1538-4357/ab08ee}

\bibitem[{{Mannheim}(1993)}]{1993A&A...269...67M}
{Mannheim}, K. 1993, \aap, 269, 67, \dodoi{10.48550/arXiv.astro-ph/9302006}

\bibitem[{{Maraschi} {et~al.}(1992){Maraschi}, {Ghisellini}, \&
  {Celotti}}]{1992ApJ...397L...5M}
{Maraschi}, L., {Ghisellini}, G., \& {Celotti}, A. 1992, \apjl, 397, L5,
  \dodoi{10.1086/186531}

\bibitem[{{Massaro} {et~al.}(2004){Massaro}, {Perri}, {Giommi}, {Nesci}, \&
  {Verrecchia}}]{2004A&A...422..103M}
{Massaro}, E., {Perri}, M., {Giommi}, P., {Nesci}, R., \& {Verrecchia}, F.
  2004, \aap, 422, 103, \dodoi{10.1051/0004-6361:20047148}

\bibitem[{{McHardy}(2010)}]{2010LNP...794..203M}
{McHardy}, I. 2010, in Lecture Notes in Physics, Berlin Springer Verlag, ed.
  T.~{Belloni}, Vol. 794, 203, \dodoi{10.1007/978-3-540-76937-8_8}

\bibitem[{{Meyer} {et~al.}(2019){Meyer}, {Scargle}, \&
  {Blandford}}]{2019ApJ...877...39M}
{Meyer}, M., {Scargle}, J.~D., \& {Blandford}, R.~D. 2019, \apj, 877, 39,
  \dodoi{10.3847/1538-4357/ab1651}

\bibitem[{{M{\"u}cke} \& {Protheroe}(2001)}]{2001APh....15..121M}
{M{\"u}cke}, A., \& {Protheroe}, R.~J. 2001, Astroparticle Physics, 15, 121,
  \dodoi{10.1016/S0927-6505(00)00141-9}

\bibitem[{{M{\"u}cke} {et~al.}(2003){M{\"u}cke}, {Protheroe}, {Engel},
  {Rachen}, \& {Stanev}}]{2003APh....18..593M}
{M{\"u}cke}, A., {Protheroe}, R.~J., {Engel}, R., {Rachen}, J.~P., \& {Stanev},
  T. 2003, Astroparticle Physics, 18, 593,
  \dodoi{10.1016/S0927-6505(02)00185-8}

\bibitem[{{Narayan} \& {Piran}(2012)}]{2012MNRAS.420..604N}
{Narayan}, R., \& {Piran}, T. 2012, \mnras, 420, 604,
  \dodoi{10.1111/j.1365-2966.2011.20069.x}

\bibitem[{{Nolan} {et~al.}(2012){Nolan}, {Abdo}, {Ackermann}, {Ajello},
  {Allafort}, {Antolini}, {Atwood}, {Axelsson}, {Baldini}, {Ballet},
  {Barbiellini}, {Bastieri}, {Bechtol}, {Belfiore}, {Bellazzini}, {Berenji},
  {Bignami}, {Blandford}, {Bloom}, {Bonamente}, {Bonnell}, {Borgland},
  {Bottacini}, {Bouvier}, {Brandt}, {Bregeon}, {Brigida}, {Bruel}, {Buehler},
  {Burnett}, {Buson}, {Caliandro}, {Cameron}, {Campana}, {Ca{\~n}adas},
  {Cannon}, {Caraveo}, {Casandjian}, {Cavazzuti}, {Ceccanti}, {Cecchi},
  {{\c{C}}elik}, {Charles}, {Chekhtman}, {Cheung}, {Chiang}, {Chipaux},
  {Ciprini}, {Claus}, {Cohen-Tanugi}, {Cominsky}, {Conrad}, {Corbet}, {Cutini},
  {D'Ammando}, {Davis}, {de Angelis}, {DeCesar}, {DeKlotz}, {De Luca}, {den
  Hartog}, {de Palma}, {Dermer}, {Digel}, {Silva}, {Drell}, {Drlica-Wagner},
  {Dubois}, {Dumora}, {Enoto}, {Escande}, {Fabiani}, {Falletti}, {Favuzzi},
  {Fegan}, {Ferrara}, {Focke}, {Fortin}, {Frailis}, {Fukazawa}, {Funk},
  {Fusco}, {Gargano}, {Gasparrini}, {Gehrels}, {Germani}, {Giebels},
  {Giglietto}, {Giommi}, {Giordano}, {Giroletti}, {Glanzman}, {Godfrey},
  {Grenier}, {Grondin}, {Grove}, {Guillemot}, {Guiriec}, {Gustafsson},
  {Hadasch}, {Hanabata}, {Harding}, {Hayashida}, {Hays}, {Hill}, {Horan},
  {Hou}, {Hughes}, {Iafrate}, {Itoh}, {J{\'o}hannesson}, {Johnson}, {Johnson},
  {Johnson}, {Johnson}, {Kamae}, {Katagiri}, {Kataoka}, {Katsuta}, {Kawai},
  {Kerr}, {Kn{\"o}dlseder}, {Kocevski}, {Kuss}, {Lande}, {Landriu},
  {Latronico}, {Lemoine-Goumard}, {Lionetto}, {Llena Garde}, {Longo},
  {Loparco}, {Lott}, {Lovellette}, {Lubrano}, {Madejski}, {Marelli}, {Massaro},
  {Mazziotta}, {McConville}, {McEnery}, {Mehault}, {Michelson}, {Minuti},
  {Mitthumsiri}, {Mizuno}, {Moiseev}, {Mongelli}, {Monte}, {Monzani},
  {Morselli}, {Moskalenko}, {Murgia}, {Nakamori}, {Naumann-Godo}, {Norris},
  {Nuss}, {Nymark}, {Ohno}, {Ohsugi}, {Okumura}, {Omodei}, {Orlando}, {Ormes},
  {Ozaki}, {Paneque}, {Panetta}, {Parent}, {Perkins}, {Pesce-Rollins},
  {Pierbattista}, {Pinchera}, {Piron}, {Pivato}, {Porter}, {Racusin},
  {Rain{\`o}}, {Rand o}, {Razzano}, {Razzaque}, {Reimer}, {Reimer}, {Reposeur},
  {Ritz}, {Rochester}, {Romani}, {Roth}, {Rousseau}, {Ryde}, {Sadrozinski},
  {Salvetti}, {Sanchez}, {Saz Parkinson}, {Sbarra}, {Scargle}, {Schalk},
  {Sgr{\`o}}, {Shaw}, {Shrader}, {Siskind}, {Smith}, {Spandre}, {Spinelli},
  {Stephens}, {Strickman}, {Suson}, {Tajima}, {Takahashi}, {Takahashi},
  {Tanaka}, {Thayer}, {Thayer}, {Thompson}, {Tibaldo}, {Tibolla}, {Tinebra},
  {Tinivella}, {Torres}, {Tosti}, {Troja}, {Uchiyama}, {Vandenbroucke}, {Van
  Etten}, {Van Klaveren}, {Vasileiou}, {Vianello}, {Vitale}, {Waite},
  {Wallace}, {Wang}, {Werner}, {Winer}, {Wood}, {Wood}, {Wood}, {Yang}, \&
  {Zimmer}}]{2012ApJS..199...31N}
{Nolan}, P.~L., {Abdo}, A.~A., {Ackermann}, M., {et~al.} 2012, \apjs, 199, 31,
  \dodoi{10.1088/0067-0049/199/2/31}

\bibitem[{{Poole} {et~al.}(2008){Poole}, {Breeveld}, {Page}, {Landsman},
  {Holland}, {Roming}, {Kuin}, {Brown}, {Gronwall}, {Hunsberger}, {Koch},
  {Mason}, {Schady}, {vanden Berk}, {Blustin}, {Boyd}, {Broos}, {Carter},
  {Chester}, {Cucchiara}, {Hancock}, {Huckle}, {Immler}, {Ivanushkina},
  {Kennedy}, {Marshall}, {Morgan}, {Pandey}, {de Pasquale}, {Smith}, \&
  {Still}}]{2008MNRAS.383..627P}
{Poole}, T.~S., {Breeveld}, A.~A., {Page}, M.~J., {et~al.} 2008, \mnras, 383,
  627, \dodoi{10.1111/j.1365-2966.2007.12563.x}

\bibitem[{{Prokhorov} \& {Moraghan}(2017)}]{2017MNRAS.471.3036P}
{Prokhorov}, D.~A., \& {Moraghan}, A. 2017, \mnras, 471, 3036,
  \dodoi{10.1093/mnras/stx1742}

\bibitem[{{Qian}(2023)}]{2023arXiv230606863Q}
{Qian}, S.~J. 2023, arXiv e-prints, arXiv:2306.06863,
  \dodoi{10.48550/arXiv.2306.06863}

\bibitem[{{Roming} {et~al.}(2005){Roming}, {Kennedy}, {Mason}, {Nousek}, {Ahr},
  {Bingham}, {Broos}, {Carter}, {Hancock}, {Huckle}, {Hunsberger}, {Kawakami},
  {Killough}, {Koch}, {McLelland}, {Smith}, {Smith}, {Soto}, {Boyd},
  {Breeveld}, {Holland}, {Ivanushkina}, {Pryzby}, {Still}, \&
  {Stock}}]{2005SSRv..120...95R}
{Roming}, P. W.~A., {Kennedy}, T.~E., {Mason}, K.~O., {et~al.} 2005, \ssr, 120,
  95, \dodoi{10.1007/s11214-005-5095-4}

\bibitem[{{Rybicki} \& {Lightman}(1986)}]{1986rpa..book.....R}
{Rybicki}, G.~B., \& {Lightman}, A.~P. 1986, {Radiative Processes in
  Astrophysics}

\bibitem[{{Sahakyan}(2018)}]{2018ApJ...866..109S}
{Sahakyan}, N. 2018, \apj, 866, 109, \dodoi{10.3847/1538-4357/aadade}

\bibitem[{{Sahayanathan}(2008)}]{2008MNRAS.388L..49S}
{Sahayanathan}, S. 2008, \mnras, 388, L49,
  \dodoi{10.1111/j.1745-3933.2008.00497.x}

\bibitem[{{Sahayanathan} {et~al.}(2018){Sahayanathan}, {Sinha}, \&
  {Misra}}]{2018RAA....18...35S}
{Sahayanathan}, S., {Sinha}, A., \& {Misra}, R. 2018, Research in Astronomy and
  Astrophysics, 18, 035, \dodoi{10.1088/1674-4527/18/3/35}

\bibitem[{{Scargle} {et~al.}(2013){Scargle}, {Norris}, {Jackson}, \&
  {Chiang}}]{2013ApJ...764..167S}
{Scargle}, J.~D., {Norris}, J.~P., {Jackson}, B., \& {Chiang}, J. 2013, \apj,
  764, 167, \dodoi{10.1088/0004-637X/764/2/167}

\bibitem[{{Schleicher} {et~al.}(2019){Schleicher}, {Arbet-Engels}, {Baack},
  {Balbo}, {Biland}, {Blank}, {Bretz}, {Bruegge}, {Bulinski}, {Buss}, {Doerr},
  {Dorner}, {Elsaesser}, {Grischagin}, {Hildebrand}, {Linhoff}, {Mannheim},
  {Mueller}, {Neise}, {Neronov}, {Noethe}, {Paravac}, {Rhode}, {Schulz},
  {Sedlaczek}, {Shukla}, {Sliusar}, {Willert}, \&
  {Walter}}]{2019Galax...7...62S}
{Schleicher}, B., {Arbet-Engels}, A., {Baack}, D., {et~al.} 2019, Galaxies, 7,
  62, \dodoi{10.3390/galaxies7020062}

\bibitem[{{Shah}(2024)}]{2024MNRAS.527.5140S}
{Shah}, Z. 2024, \mnras, 527, 5140, \dodoi{10.1093/mnras/stad3534}

\bibitem[{{Shah} {et~al.}(2019){Shah}, {Jithesh}, {Sahayanathan}, {Misra}, \&
  {Iqbal}}]{2019MNRAS.484.3168S}
{Shah}, Z., {Jithesh}, V., {Sahayanathan}, S., {Misra}, R., \& {Iqbal}, N.
  2019, \mnras, 484, 3168, \dodoi{10.1093/mnras/stz151}

\bibitem[{{Shah} {et~al.}(2018){Shah}, {Mankuzhiyil}, {Sinha}, {Misra},
  {Sahayanathan}, \& {Iqbal}}]{2018RAA....18..141S}
{Shah}, Z., {Mankuzhiyil}, N., {Sinha}, A., {et~al.} 2018, Research in
  Astronomy and Astrophysics, 18, 141, \dodoi{10.1088/1674-4527/18/11/141}

\bibitem[{{Shah} {et~al.}(2017){Shah}, {Sahayanathan}, {Mankuzhiyil},
  {Kushwaha}, {Misra}, \& {Iqbal}}]{2017MNRAS.470.3283S}
{Shah}, Z., {Sahayanathan}, S., {Mankuzhiyil}, N., {et~al.} 2017, \mnras, 470,
  3283, \dodoi{10.1093/mnras/stx1194}

\bibitem[{{Sikora} {et~al.}(1994){Sikora}, {Begelman}, \&
  {Rees}}]{1994ApJ...421..153S}
{Sikora}, M., {Begelman}, M.~C., \& {Rees}, M.~J. 1994, \apj, 421, 153,
  \dodoi{10.1086/173633}

\bibitem[{{Sinha} {et~al.}(2018){Sinha}, {Khatoon}, {Misra}, {Sahayanathan},
  {Mandal}, {Gogoi}, \& {Bhatt}}]{2018MNRAS.480L.116S}
{Sinha}, A., {Khatoon}, R., {Misra}, R., {et~al.} 2018, \mnras, 480, L116,
  \dodoi{10.1093/mnrasl/sly136}

\bibitem[{{Tantry} {et~al.}(2024){Tantry}, {Shah}, {Misra}, {Iqbal}, \&
  {Akbar}}]{2024arXiv241100592T}
{Tantry}, J., {Shah}, Z., {Misra}, R., {Iqbal}, N., \& {Akbar}, S. 2024, arXiv
  e-prints, arXiv:2411.00592.
\newblock \doarXiv{2411.00592}

\bibitem[{Tripathi {et~al.}(2023)Tripathi, Gupta, Takey, Bachev, Vince,
  Strigachev, Kushwaha, Elhosseiny, Wiita, Damljanovic, Dhiman, Fouad, Gaur,
  Gu, Hamed, Kishore, Kurtenkov, Rastogi, Semkov, Zead, \&
  Zhang}]{10.1093/mnras/stad3574}
Tripathi, T., Gupta, A.~C., Takey, A., {et~al.} 2023, Monthly Notices of the
  Royal Astronomical Society, 527, 5220, \dodoi{10.1093/mnras/stad3574}

\bibitem[{{Urry} \& {Padovani}(1995)}]{1995PASP..107..803U}
{Urry}, C.~M., \& {Padovani}, P. 1995, \pasp, 107, 803, \dodoi{10.1086/133630}

\bibitem[{{Vaughan} {et~al.}(2003){Vaughan}, {Edelson}, {Warwick}, \&
  {Uttley}}]{2003MNRAS.345.1271V}
{Vaughan}, S., {Edelson}, R., {Warwick}, R.~S., \& {Uttley}, P. 2003, \mnras,
  345, 1271, \dodoi{10.1046/j.1365-2966.2003.07042.x}

\bibitem[{{Villata} {et~al.}(2006){Villata}, {Raiteri}, {Balonek}, {Aller},
  {Jorstad}, {Kurtanidze}, {Nicastro}, {Nilsson}, {Aller}, {Arai}, {Arkharov},
  {Bach}, {Ben{\'\i}tez}, {Berdyugin}, {Buemi}, {B{\"o}ttcher}, {Carosati},
  {Casas}, {Caulet}, {Chen}, {Chiang}, {Chou}, {Ciprini}, {Coloma}, {di Rico},
  {D{\'\i}az}, {Efimova}, {Forsyth}, {Frasca}, {Fuhrmann}, {Gadway}, {Gupta},
  {Hagen-Thorn}, {Harvey}, {Heidt}, {Hernandez-Toledo}, {Hroch}, {Hu}, {Hudec},
  {Ibrahimov}, {Imada}, {Kamata}, {Kato}, {Katsuura}, {Konstantinova},
  {Kopatskaya}, {Kotaka}, {Kovalev}, {Kovalev}, {Krichbaum}, {Kubota},
  {Kurosaki}, {Lanteri}, {Larionov}, {Larionova}, {Laurikainen}, {Lee}, {Leto},
  {L{\"a}hteenm{\"a}ki}, {L{\'o}pez-Cruz}, {Marilli}, {Marscher}, {McHardy},
  {Mondal}, {Mullan}, {Napoleone}, {Nikolashvili}, {Ohlert}, {Postnikov},
  {Pursimo}, {Ragni}, {Ros}, {Sadakane}, {Sadun}, {Savolainen}, {Sergeeva},
  {Sigua}, {Sillanp{\"a}{\"a}}, {Sixtova}, {Sumitomo}, {Takalo},
  {Ter{\"a}sranta}, {Tornikoski}, {Trigilio}, {Umana}, {Volvach}, {Voss}, \&
  {Wortel}}]{2006A&A...453..817V}
{Villata}, M., {Raiteri}, C.~M., {Balonek}, T.~J., {et~al.} 2006, \aap, 453,
  817, \dodoi{10.1051/0004-6361:20064817}

\bibitem[{{White} {et~al.}(1988){White}, {Jauncey}, {Savage}, {Wright},
  {Batty}, {Peterson}, \& {Gulkis}}]{1988ApJ...327..561W}
{White}, G.~L., {Jauncey}, D.~L., {Savage}, A., {et~al.} 1988, \apj, 327, 561,
  \dodoi{10.1086/166216}

\bibitem[{{Wood} {et~al.}(2017){Wood}, {Caputo}, {Charles}, {Di Mauro},
  {Magill}, {Perkins}, \& {Fermi-LAT Collaboration}}]{2017ICRC...35..824W}
{Wood}, M., {Caputo}, R., {Charles}, E., {et~al.} 2017, in International Cosmic
  Ray Conference, Vol. 301, 35th International Cosmic Ray Conference
  (ICRC2017), 824, \dodoi{10.22323/1.301.0824}

\bibitem[{Xiao {et~al.}(2022)Xiao, Ouyang, Zhang, Fu, Zhang, Zeng, \&
  Fan}]{Xiao_2022}
Xiao, H., Ouyang, Z., Zhang, L., {et~al.} 2022, The Astrophysical Journal, 925,
  40, \dodoi{10.3847/1538-4357/ac36da}

\bibitem[{{Zirakashvili} \& {Aharonian}(2007)}]{2007A&A...465..695Z}
{Zirakashvili}, V.~N., \& {Aharonian}, F. 2007, \aap, 465, 695,
  \dodoi{10.1051/0004-6361:20066494}

\end{thebibliography}
\bibliographystyle{aasjournal}



\end{document}